\documentclass[journal]{IEEEtran}
\ifCLASSINFOpdf
\else
\fi
\usepackage{hyperref}
\usepackage[utf8]{inputenc}
\usepackage{amssymb}
\usepackage{bbm}
\usepackage{amsfonts}
\usepackage[dvips]{graphicx}
\usepackage{times}
\usepackage[]{cite}
\usepackage{amsmath}
\usepackage{array}
\usepackage{amssymb}

\usepackage{stfloats}
\usepackage{slashbox}
\usepackage{graphicx}
\usepackage{footnote}
\usepackage{booktabs}
\usepackage{array}
\usepackage{mathtools}
\usepackage{dsfont}
\usepackage{ntheorem}
\usepackage{subeqnarray}
\usepackage{cases}
\usepackage{threeparttable}
\usepackage{color}
\usepackage{nomencl}
\makenomenclature
\usepackage[numbers,sort&compress]{natbib}
\IEEEoverridecommandlockouts
\newtheorem{assumption}{Assumption}[section]
\newtheorem{theorem}{Theorem}

\newtheorem{lemma}{Lemma}

\usepackage{flafter}
\usepackage{subfigure}
\usepackage{float}
\usepackage{bbold}
\usepackage{supertabular}
\usepackage{caption}
\usepackage{url}
\usepackage{xcolor}
\usepackage[ruled,vlined,lined,ruled,linesnumbered]{algorithm2e}
\newtheorem{definition}{Definition}
\newcommand{\bm}[1]{\mbox{\boldmath{$#1$}}}

\DeclareMathOperator*{\argmin}{arg\,min}
\usepackage{subfigure}
\usepackage{stfloats}
\usepackage{etoolbox}
\renewcommand\nomgroup[1]{%
  \item[\bfseries
  \ifstrequal{#1}{P}{Physics Constants}{%
  \ifstrequal{#1}{N}{Number Sets}{%
  \ifstrequal{#1}{O}{Other Symbols}{}}}%
]}
\allowdisplaybreaks
\SetKwProg{Fn}{Function}{}{end}\SetKwFunction{FRecurs}{FnRecursive}%

\begin{document}
%
\title{Constrained Reinforcement Learning for Stochastic Dynamic  Optimal Power Flow Control}
\allowdisplaybreaks
\IEEEaftertitletext{\vspace{-2.2\baselineskip}}

\author{Tong~Wu,~\IEEEmembership{Member,~IEEE,}
        Anna~Scaglione,~\IEEEmembership{Fellow,~IEEE,}, \IEEEauthorblockN{Daniel Arnold,~\IEEEmembership{Member,~IEEE}}
\thanks{Tong Wu and Anna Scaglione   are with the Department of Electrical and Computer Engineering, Cornell Tech, Cornell University, 10044 USA (e-mail: \{tw385, as337\}@cornell.edu). Daniel Arnold is with Lawrence Berkeley National Laboratory (e-mail: dbarnold@lbl.gov).  This research was supported in part    by the National Science Foundation (NSF) under Grant NSF ECCS \# 2210012; and in part by the Director, Cybersecurity, Energy Security, and Emergency Response, Cybersecurity for Energy Delivery Systems program, of the U.S. Department of Energy, under contract DE-AC02-05CH11231. Any opinions, findings, conclusions, or recommendations expressed in this material are those of the authors and do not necessarily reflect those of the sponsors of this work.}
}


\newcommand{\norm}[1]{\left\lVert#1\right\rVert}
\newcommand*\abs[1]{\lvert#1\rvert}

\IEEEtitleabstractindextext{%
\begin{abstract}
 Deep Reinforcement Learning (DRL) has become a popular method for solving control problems in power systems. Conventional DRL encourages the agent to explore various policies encoded in a neural network (NN) with the goal of maximizing the reward function. However, this approach can lead to infeasible solutions that violate physical constraints such as power flow equations, voltage limits, and dynamic constraints. Ensuring these constraints are met is crucial in power systems, as they are a safety critical infrastructure. To address this issue, existing DRL algorithms remedy the problem by projecting the actions onto the feasible set, which can result in sub-optimal solutions. This paper presents a novel primal-dual approach for learning optimal constrained DRL policies for dynamic optimal power flow problems, with the aim of controlling power generations and battery outputs. We also prove the convergence of the critic and actor networks. Our case studies on IEEE standard systems demonstrate the superiority of the proposed approach in dynamically adapting to the environment while maintaining safety constraints.
\end{abstract}

}

\maketitle

\IEEEdisplaynontitleabstractindextext

%
\IEEEpeerreviewmaketitle

\section*{Nomenclature}
\vspace{-0.4cm}
\textit{Sets}
\addcontentsline{toc}{section}{Nomenclature}
\begin{IEEEdescription}[\IEEEsetlabelwidth{$V_1,V_2$}]\small
\item[$\mathcal{N}$] The set of all buses with cardinality $N$.
\item[$\mathcal{G}$] The set of all  buses with generators installed (The number of generations is $G$).
\item[$\mathcal{G}_s$ ($\mathcal{G}_n$)] The set  of (non-) slack buses with generation installed.
\item[$\mathcal{B}$] The set of all   BESSs with cardinality $B$.
\end{IEEEdescription}

\textit{Variables}
\addcontentsline{toc}{section}{Nomenclature}
\begin{IEEEdescription}[\IEEEsetlabelwidth{$V_1,V_2,V_3,V_4$}]\small
\item[$\bm{g}^p_t(\hat{\bm{g}}^p_t)$] The vector of  active generations (normalized generations in $[0, 1]^{G}$) at time $t$.
\item[$\bm{g}^q_t(\hat{\bm{g}}^q_t)$] The vector of  reactive generations (normalized generations in $[0, 1]^{G}$) at time $t$.
\item[$\bm{p}_{dis, t}(\hat{\bm{p}}_{dis, t})$] The vector of  discharging powers of BESSs (normalized versions in $[0, 1]^{B}$) at time $t$.
\item[$\bm{p}_{ch, t}(\hat{\bm{p}}_{ch, t})$] The vector of   charging powers of BESSs (normalized versions in $[0, 1]^{B}$) at time $t$.
\item[$\bm{v}_t$] The vector of voltage phasors at time $t$.
\end{IEEEdescription}

\textit{Constants}
\addcontentsline{toc}{section}{Nomenclature}
\begin{IEEEdescription}[\IEEEsetlabelwidth{$V_1,V_2,V_3,V_4v$}]\small
\item[$\Delta t$] Length of time step.
\item[$E_{cap}$] Energy capacity of BESS.
\item[$\eta_{ch}$] Charging efficiency.
\item[$\eta_{dis}$] Discharging efficiency.
\item[$\bm{P}^{ch}_{rated}$] The vector of maximal charging powers.
\item[$\bm{P}^{dis}_{rated}$] The vector of maximal discharging powers.
\item[$\bm{d}^p_t$] The vector of active demands at time $t$.
\item[$\bm{d}^q_t$] The vector of active demands at time $t$.
\item[$\mathbf{Y}$] The admittance matrix.
\item[$\bm{soc}_{\min}$($\bm{soc}_{\max}$)] The vector of minimal (maximal)  charge limits.
\item[$\mathbf{M}_{g}$] The mapping matrix $\{0,1\}^{N\times G}$ to expend $\bm{g}_t$ from $\mathbb{R}^{G}$ to $\mathbb{R}^{N}$, where  $[\mathbf{M}_{g}\bm{g}^p_{t}]_{i}$  is 0,  $\forall i\in \mathcal{N}/\mathcal{G}$, and otherwise is ${g}_{i,t}, \forall i\in  \mathcal{G}$
\item[$\mathbf{M}_{b}$] The mapping matrix $\{0,1\}^{N\times B}$ to expend $\bm{p}_{dis,t}$   from $\mathbb{R}^{B}$ to $\mathbb{R}^{N}$, where  $[\mathbf{M}_{b}\bm{p}_{dis,t}]_{i}$  is 0,  $\forall i\in \mathcal{N}/\mathcal{B}$, and otherwise is ${p}_{dis,i, t}, \forall i\in  \mathcal{B}$
\end{IEEEdescription}

\ifCLASSOPTIONcompsoc
\IEEEraisesectionheading{\section{Introduction}\label{sec:introduction}}
\else
\section{Introduction}
\label{sec:introduction}
\fi

\subsection{Background and Motivation}
The power grid is a complex, dynamic network composed of interconnected components that can be influenced by numerous factors, including fluctuations in demand, changes in energy resource availability, and the operation of power plants and control assets (e.g. frequency control and voltage regulation) \cite{machowski2020power}. The increasing penetration of renewable and decentralized energy resources (DER) poses significant operational challenges for power networks operators, because of the need to manage their dynamic behavior. At the same time, the widespread deployment of advanced measurement technologies such as Phasor Measurements Units  (PMUs) in the bulk system, and the Advanced Metering Infrastructure (AMI), in distribution systems, provides new opportunities to leverage the data for real-time power network control  \cite{chen2022reinforcement}, rather than relying only on local control loops to respond to the grid state.

From an operational perspective, in the presence of the uncertainty not only of demand but also of DER generation, the challenge of optimal control of dynamic devices  such as battery energy storage systems (BESSs), is being addressed through the formulation of stochastic dynamic optimal power flow (SDOPF) methods. These methods dispatch generation resources and select BESSs charging or discharging periods accounting for the future impact of real-time decision-making, to ensure efficient and reliable operations \cite{gill2013dynamic}. In fact, a SDOPF formulation solves  the general problem of how to optimally dispatch generation and operating storage units across a network to meet net electric load within a time-horizon economically, accounting for the dynamic constraints of the electric power supply sources \cite{guo2018data}.  However, its implementation in real-time is challenging due to the unpredictable nature of DER and demand of electric power, the dynamic constraints of generation and storage, and the computational complexity of the SDOPF problem.

Deep reinforcement learning (DRL) has gained significant attention for its potential to learn SDOPF policies in dynamic grid control applications, such as BESS management and EV charging control with dynamic constraints. These studies aim to derive optimal policies for decision making under uncertainty by training the algorithm offline on real-world scenarios \cite{bui2019double, cao2020deep, gorostiza2020deep, al2020reinforcement, chen2022reinforcement}. DRL offers a promising solution for optimizing the real-time operation of BESSs by training offline on real scenarios to handle uncertainty  \cite{chen2022reinforcement}.
Examples of DRL applications for BESS management include a distributed operation strategy using double deep Q-learning for community BESS in microgrids \cite{bui2019double}, a noisy network-based approach for optimizing charging/discharging \cite{cao2020deep}, a controller to manage the  state of charge (SOC) of multiple BESSs providing frequency support to the grid \cite{gorostiza2020deep}, and a Monte Carlo tree search-based approach to alleviate the BESS capacity problem in a cooperative  SOC   control scheme \cite{al2020reinforcement}. In \cite{ding2020optimal},  an optimal strategy for electric vehicle (EV) charging was developed in a distribution network through the use of reinforcement learning, taking into account the dynamic constraints of the  state of charge (SOC)  of the vehicles.

\subsection{Related Works}
In our review of the prior art we will focus on stochastic dynamic optimal power flow (OPF) and learning-methods to solve OPF formulations, as well as the literature on constrained reinforcement learning.

\subsubsection{Stochastic Dynamic OPF}  
A stochastic OPF formulation was first introduced in \cite{yong2000stochastic} to solve the optimal dispatch problem with uncertainties in power systems. 
\cite{bazrafshan2016decentralized} proposed a   stochastic OPF method for radial distribution networks with high levels of PV penetration.   \cite{guo2018data} introduced a data-driven approach to solve multi-stage stochastic OPF with limited information on forecast errors, providing closed-loop control policies.  In \cite{usman2022three}, a multi-stage OPF formulation with BESS was studied and three novel optimization methods were proposed to solve multi-period ACOPF.  However, these iterative optimization methods have very high computational complexity, limiting their promise for real-time control.

\subsubsection{Learning-based OPF}  Recently, learning-based approaches for solving OPF problems swiftly have received substantial attention. In a nutshell, the  idea behind these algorithms  is to leverage the universal approximation   capabilities of DNNs to learn the mapping between load input and OPF solutions \cite{pan2022deepopf, singh2021learning, huang2021deepopf, wu2021deep}. Then one can pass as input to the trained DNN the network load and instantly obtain a quality solution. A key difficulty for applying DNN to solve AC-OPF problems lies in   that the solutions may not satisfy the physical and operational constraints that make the solution feasible. To address this problem, \cite{pan2022deepopf} includes a  regularizaqtion term in the DNN reward objective that penalizes solutions that are AC-OPF infeasible. In \cite{wu2021deep}, instead, a small-scale mapping method was proposed to recover the feasible results. The unsupervised idea is to learn the solution in an unsupervised manner, minimizing the cost directly  \cite{donti2020dc3, owerko2022unsupervised}.   \cite{donti2020dc3} considered both the penalty function and mapping function for both equality and inequality constraints.   In \cite{owerko2022unsupervised},  piece-wise penalty function based on the log-barrier is considered  to enforce constraints. However, these learning-based methods cannot consider the dynamic constraints and how the current actions affect the future.

\subsubsection{Constrained Reinforcement Learning}  
Deep reinforcement learning methods are promising when solving complex stochastic nonlinear dynamic control problems that look at  maximizing not just current but also future rewards from the control action. However, as mentioned before, the  DRL policies may produce decisions that are infeasible as they  violate the power flow equations and SOC limits. To avoid this issue Constrained Reinforcement Learning (CRL) has taken front-stage for solving constrained sequential decision-making problems in safety critical systems. The Lagrangian relaxation is one of the effective approaches to address CRL problems (see a review \cite{liu2021policy}) \cite{achiam2017constrained, liu2020ipo, chow2017risk}. 
In addition to the paper cited above, \cite{ding2022convergence, ding2020natural} follow a Natural Policy Gradient Primal-Dual, giving guarantees for convergence to a fixed point. \cite{paternain2019learning} considers the chance-constrained reinforcement learning by primal-dual methods. \cite{qiu2020upper} provides a upper confidence primal-dual algorithm and proves upper-bounds of both the regret and the constraint violation.  Similar work  on CRL application in OPF consider that operational constraints are satisfied by a novel convex safety layer based on the penalty convex-concave procedure \cite{sayed2022feasibility}. In \cite{yan2020real}, a Lagrangian based DRL is considered to optimize OPF function. However, this method is hard to scale to multi-stage dynamic constraints because its design is too simple to handle equation constraints.

\subsection{Contributions and Organization}
The CRL methods cited above mainly consider aggregate constraints, requiring that the sum of one constrained variable from the beginning to the current time step are bounded within a certain limit.  In contrast, in the multi-stage stochastic   dynamic OPF,  the power-flow and dynamic constraints need to be met at each time step. The main contribution of this paper is summarized as follows:
\begin{itemize}
    \item We propose a training framework for  CRL that ensures the actions selected by the policy are feasible at each time step. Specifically, we modify the twin delayed deep deterministic policy gradient algorithm (TD3) \cite{fujimoto2018addressing} to optimize the control of power generation and  BESS  charging and discharging actions in a multi-stage SDOPF problem.
 \item We use the augmented Lagrangian method to solve the constrained SDOPF and update the dual variables of the modified TD3 using primal-dual methods.
 \item We introduce a complex-valued graph convolutional network for the actor to capture the spatiotemporal correlation of the environment states.
 \item We prove the convergence of critic networks and, under mild assumptions, the convergence of the augmented Lagrangian actor networks.
\end{itemize}

The rest of the paper is organized as follows. Section II presents  the SDOPF problem. Section III proposed the constrained reinforcement learning method. Then, in Section IV, we introduce the complex-valued graph convolutional policy function. We further provide the convergence  analysis of the constrained reinforcement learning method in Section V. We then implement the proposed primal-dual constrained reinforcement learning method for a specific case study in Section VI.  Experimental simulations are carried out to validate the effectiveness the proposed approach in Section VII. Finally, Section VIII draws some conclusions. 

\section{Problem Formulation}
The problem solved in this paper is an instance of the following a multi-stage stochastic optimal control formulation:
\begin{subequations}
\begin{alignat}{2}
	\min_{   \pi(\boldsymbol{a}_t|\boldsymbol{x}_{t-1})}   \mathbb{E}_{\boldsymbol{d},\pi(\cdot)}\left[\sum_{t = \tau}^{\tau+T-1} \ell_t(\bm{x}_t, \bm{a}_{t}, \bm{d}_t)\right]\label{eqobj1} \\
	   \bm{x}_{t} = f_t(\bm{x}_{t-1}, \bm{a}_{t-1},\bm{d}_{t-1}),\label{eqconstr1}\\
      \bm{a}_t \sim \pi(\bm a_t|\bm x_{t-1}),~~(\bm{x}_t, \bm{a}_t) \in \chi_t, \label{eqconstr3}
\end{alignat}
\end{subequations}
where $\bm{x}_t$ denotes a state vector at time $t$, $\bm{a}_t$ denotes a control vector at time $t$ that includes all controllable devices in power grids, $\bm{d}_t$ denotes a random  vector that includes forecast errors and uncertainties, $\ell_t(\bm{x}_t, \bm{a}_{t}, \bm{d}_t)$ represents the cost function, $\chi_t$ represents network and device bound constraints, the system dynamics function $f_t$ models internal dynamics and other temporal interdependencies of devices, such as  SOC  for BESS, and $\pi(\bm a_t|\bm x_t)$ is the randomized policy. 
 
The stochastic SDOPF is a standard multi-stage stochastic optimal control problem to achieve the economic dispatch of power flows by controlling power generations and BESSs. In particular,  $\bm{x}_t = [\bm{v}_t; \bm{soc}_t ]^\top$ includes voltage angles $\bm{v}_t$ and the vector of SOCs $\bm{soc}_t$ of all batteries in the system. Note that $\forall i\in \mathcal{N}/\mathcal{B}, [\bm{soc}_t]_i= 0$ and $\forall i \in \mathcal{B}$, $[\bm{soc}_t]_i$ is the  state of charge of that BESS, and thus $\bm{soc}_t$ has the same dimention with $\bm{v}_t$.  The control vector $\bm{a}_t = [{\bm{g}}^p_t; \bm{g}^q_t;  {\bm{p}}_{ch,t};  {\bm{p}}_{dis,t}]^\top$  includes active power generation ${\bm{g}}^p_t = [{g}^p_{1,t}, \cdots, {g}^p_{G,t}]^\top$, reactive power generations ${\bm{g}}^q_t = [{g}^q_{1,t}, \cdots, {g}^q_{G,t}]^\top$
and ${\bm{p}}_{dis,t} = [{p}_{dis,1,t}, \cdots, {p}_{dis,B,t}]^\top$ and ${\bm{p}}_{ch,t} = [{p}_{ch,1,t}, \cdots, {p}_{ch,B,t}]^\top $ represent the charge and discharge rates of the BESSs.

\subsection{Objectives}
%
The objectives of the SDOPF with BESSs include $f_{ess}$ and $f_{cost}$.  $f_{cost}$ denotes the fuel costs as follows:
\begin{equation}
\begin{aligned}
f_{cost, t} = \sum_{i\in \mathcal{G}}( a_i g^2_{i, t} + b_i g_{i, t} + c_i  ),
\end{aligned}
\end{equation}
where $a_i$, $b_i$ and $c_i$ are positive.
$f_{ess}$ is the power loss due to charging and discharging the batteries.
\begin{equation}
\begin{aligned}
	f_{ess, t} = \sum_{i\in \mathcal{B}} (1 - \eta_{ch,i}) p_{ch,i,t} +  \left( \frac{1}{\eta_{dis,i}} -1\right) p_{dis,i,t}
\end{aligned}
\end{equation}
where $\eta_{ch,i}$ and $\eta_{dis,i}$ are the charging and discharging efficiency, respectively. $p_{ch,i,t}$ and $p_{dis,i,t}$ represent the charging and discharging power of the $i^{th}$ BESS.
The  SOC 
for each period is:
\begin{equation}
\begin{aligned}
 	soc_{i, t} = & soc_{i, t-1} + \frac{\Delta t}{E_{cap}} (        \eta_{ch}~ p_{ch,i, t}    -    \frac{1}{ \eta_{dis}}~ p_{dis,i, t}), i\in \mathcal{B}
\end{aligned}
\end{equation}
where $\Delta t$ is the duration of each decision period and $E_{cap}$ is the BESS energy capacity.
The real power of $p_{dis, i, t}$ and $p_{ch, i, t}$ are constrained as follows:
\begin{equation}
\begin{aligned}
	0\le p_{ch,i, t} \le P^{ch}_{rated}, i\in \mathcal{B}\\
	0\le p_{dis, i, t} \le  P^{dis}_{rated}, i\in \mathcal{B}
\end{aligned}
\end{equation}
where $P^{ch}_{rated}$  and $P^{dis}_{rated}$ are   charging and discharging rates limits.
The reward for action is the complement of the objectives in \eqref{eqobj1} which we want to minimize: 
\begin{equation}
\begin{aligned}\label{rwd1}
r_t = \sum_{t = \tau}^{\tau+T-1} -\ell_t(\bm{x}_t, \bm{a}_{t}, \xi_t) =  \sum_{t = \tau}^{\tau+T-1} \Big(-   f_{cost,t} -  f_{ess, t}\Big).
\end{aligned}
\end{equation}

\subsection{Stochastic Dynamic OPF}
In this paper we use the AC power flow  to enforce the power-flow constraints. The multi-stage SDOPF problem formulation is: 
\begin{subequations} \label{DDCOPF_all}
\begin{alignat}{2}\label{DDCOPF}
	&\min_{\pi(\boldsymbol{a}_{t}|\boldsymbol{v}_{t-1},\boldsymbol{soc}_{t-1})}   \mathbb{E}_{\boldsymbol{d}} \Big[\sum_{t = \tau}^{\tau+T-1} \ell_t(\bm{x}_t, \bm{a}_{t}, \bm{d}_t)\Big]   \\
	& \mathbf{M}_{b}\bm{p}_{dis, t}  - \mathbf{M}_{b}\bm{p}_{ch,t}  +  \mathbf{M}_{g} \bm{g}_t^p - \bm{d}^p_t = \Re\{D(\bm{v}_t \bm{v}_t^H \mathbf{Y}^H)\}, \label{DDCOPF_cst1}\\
	&  \mathbf{M}_{g} \bm{g}_t^q - \bm{d}^q_t = \Im\{D(\bm{v}_t \bm{v}_t^H \mathbf{Y}^H)\}, \label{DDCOPF_cst2}\\
	& \underline{\bm{g}}^p \le \bm{g}^p_t \le \overline{\bm{g}}^p,~~~  \underline{\bm{g}}^q \le \bm{g}^q_t \le \overline{\bm{g}}^q,~~~ \underline{\bm{v}} \le \abs{\bm{v}} \le \overline{\bm{v}} \\
&	0\le \bm{p}_{ch, t} \le \bm{P}^{ch}_{rated},  ~~~	0\le \bm{p}_{dis, t} \le \bm{P}^{dis}_{rated},  \\
& \bm{soc}_{min} \le \bm{soc}_t\le \bm{soc}_{max}, \forall t \in [\tau, \tau+T-1]\label{DDCOPF_cst6}\\
 &	\bm{soc}_t =   \bm{soc}_{t-1} + \frac{\Delta t}{E_{cap}}  \Big(\eta_{ch}  \bm{p}_{ch, t} - \frac{\bm{p}_{dis, t}}{ \eta_{dis}}  \Big),\label{DDCOPF_cst7} 
\end{alignat}
\end{subequations}
where  the active power demand vector is $\bm{d}^p_t = [{d}^p_{1,t}, \cdots, {d}^p_{N,t}]^\top$, the reactive power demand vector is  $\bm{d}^q_t = [{d}^q_{1,t}, \cdots, {d}^q_{N,t}]^\top$, $\mathbf{Y}$ is the admittance matrix and $\bm{v}_t = [{v}_{1, t}, \cdots, {v}_{N,t}]^\top$ is the grid state in the AC power flow, i.e. $\bm{v}_t  = \abs{\bm{v}_{t}} \odot e^{\mathfrak{j}  \boldsymbol{\theta}_{t} },~~{v}_{n,t} = \abs{{v}_{n,t}} e^{\mathfrak{j}  {\theta}_{n,t} } $. Let $\mathbf{M}_{g}$  the matrix $\{0,1\}^{N\times G}$ that maps the generation vector $\bm{g}^p_t\in \mathbb{R}^{|{\cal G}|}$  to $\mathbb{R}^{N}$ as follows:
\begin{equation}
\begin{aligned}
   & [\mathbf{M}_{g} \bm{g}^p]_i = 0, [\mathbf{M}_{g} \bm{g}^q]_i = 0, \quad \forall i \in \mathcal{N} \setminus \mathcal{G}\\
   & [\mathbf{M}_{g} \bm{g}^p]_i = {g}^p_j, [\mathbf{M}_{g} \bm{g}^q]_i = {g}^q_j,  \quad \forall i \in   \mathcal{G}, ~~\forall j \in [1, \cdots, G]
\end{aligned}
\end{equation}
and similarly $\mathbf{M}_{b}$ the matrix that maps the vectors $\bm{p}_{ch, t}$ and $\bm{p}_{dis, t}$ onto the entire network, adding zero in the buses that do not have batteries.
The feasible set of \eqref{DDCOPF_cst1} - \eqref{DDCOPF_cst6} is denoted by $\chi_t$. 
The method proposed to solve \eqref{DDCOPF_all} is detailed in next. 

\section{Constrained Reinforcement Learning}
In this section, we advance actor-critic policy gradient methods to incorporate the instantaneous constraints in \eqref{DDCOPF_all}.

\begin{figure}[!htb]
	\centering
	\includegraphics[width=0.8\columnwidth]{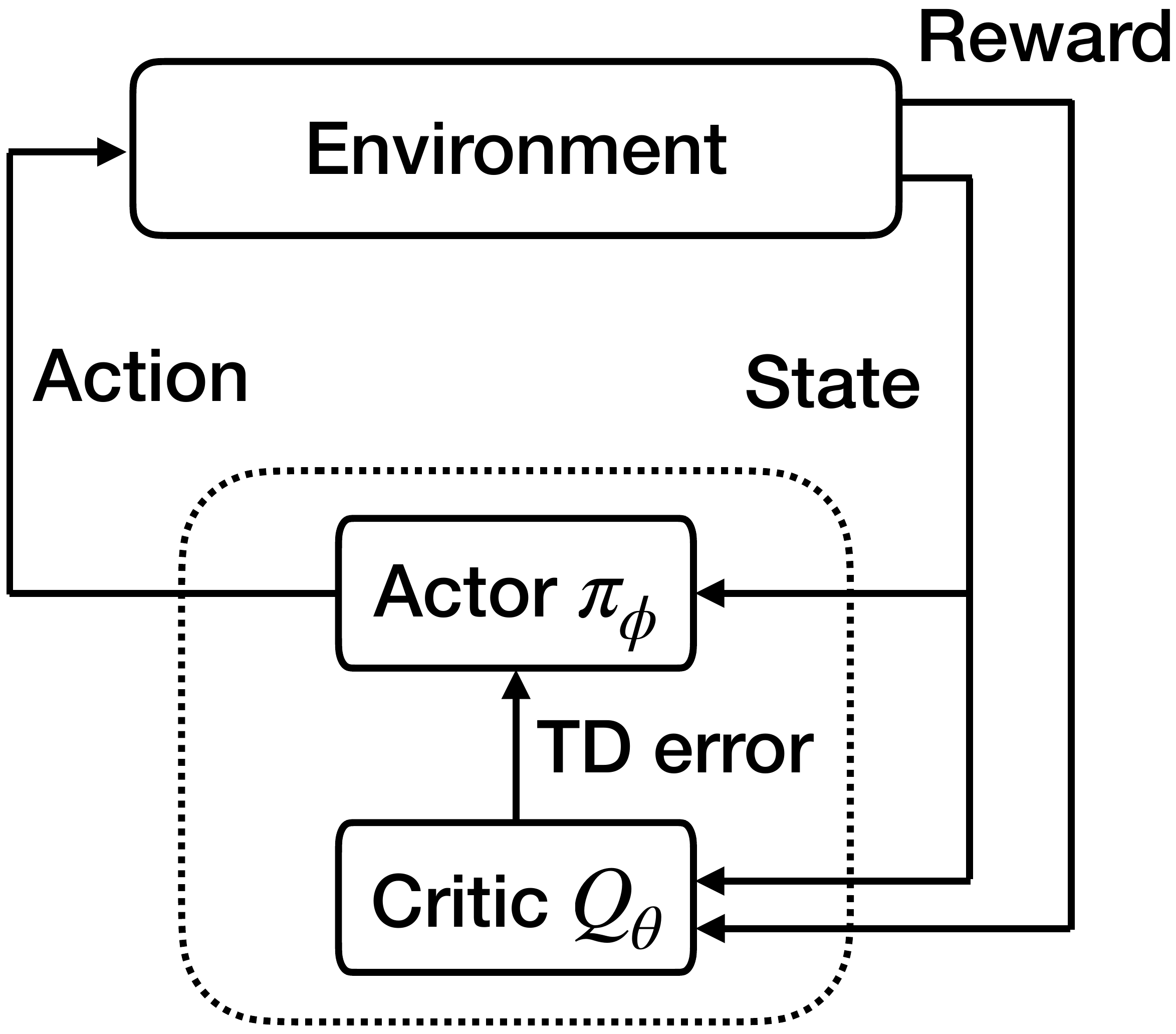}
	\caption{The actor-critic architecture.}
	\label{fig:actor-critic}
		\vspace{-0.6cm}
\end{figure}

\subsection{Actor-Critic Method}
The basic policy gradient method is an actor-only method where the actor refers to the policy function. Usually, in learning the parameters for the approximated policy function, actor-only methods are plagued by high-variance and slow learning \cite{sutton2018reinforcement}. Actor-critic method overcomes these shortcomings by updating policy function parameters based on the approximate value function, which is referred to as {\it the critic}. As shown in Fig. \ref{fig:actor-critic},  the actor is a policy function $\pi_{\phi}$ parameterized by $\phi$ for action selection, while the critic is a state-value function $Q_{\xi}$ parameterized by $\xi$ to criticize the action made by the actor. 

\subsubsection{Forecasting Action}
The tuple of actions  for the multi-stage SDOPF at time $t$ is denoted by:
\begin{equation}\label{eqobj}
\begin{split}
&\hat{\bm{a}}_t = [\hat{\bm{g}}^p_t; \hat{\bm{g}}^q_t; \hat{\bm{p}} _{ch,t}; \hat{\bm{p}}_{dis,t}]^\top,\\
& \bm{A}_{t} = [\hat{\bm{a}}_t, \cdots, \hat{\bm{a}}_{t+T-1}]^\top,
\end{split}
\end{equation}
where $\hat{\bm{a}}_t$ denotes the normalized control action by NNs, i.e. $\hat{\bm{g}}^p_t, \hat{\bm{g}}^q_t, \hat{\bm{p}}_{ch,t}, \hat{\bm{p}}_{dis,t}$ as the normalized versions of  ${\bm{g}}^p_t, {\bm{g}}^q_t, {\bm{p}}_{ch,t}, {\bm{p}}_{dis,t}$ to $[0, 1]$ (due to the sigmoid activation of NN), and $\bm{A}_{t}$ denotes the vector of control actions, i.e., $ \hat{\bm{a}}_t, \cdots,  \hat{\bm{a}}_{t+T-1}$,  over the future horizon. Notice that our control policy works on a sliding window where of all actions predicted in $\bm{A}_{t}$, only $\hat{\bm{a}}_t$ is applied to the environment. This is because $\bm{A}_{t}$ satisfying the constraints $\chi_t$ can ensure that the current action $\hat{\bm{a}}_t$ will not lead to an infeasible solution to future actions, i.e., $\hat{\bm{a}}_{t+1} \cdots, \hat{\bm{a}}_{t+T-1}$.

\subsubsection{Voltage Magnitudes}
The policy controls the future power injections, which are implicitly related to the voltage magnitudes through the power flow equations.  In order to consider the constraint $\underline{\bm{v}} \le \abs{\bm{v}} \le \overline{\bm{v}}$,  we utilize an independent  neural network to solve for the voltage magnitudes for a given action to ensure that they are are equal to the ground-truth and within the bound $[\underline{\bm{v}}, \overline{\bm{v}}]$. We define the prediction network as$\abs{\hat{\bm{v}}} = P_\omega(\bm{x})$, where $\abs{\hat{\bm{v}}}$ is defined as the normalized versions of $\abs{{\bm{v}}}$ in the range $[0, 1]$.
\subsubsection{Critic Design}
In Q-learning, the value function can be learnt by temporal difference learning \cite{sutton1988learning} based on the Bellman equation \cite{bellman1966dynamic}. The Bellman equation is a fundamental relationship between  the value of a state-action pair $(\bm{x}, \bm{A})$ and the value of the subsequent (future) state-action pair $(\bm{x}', \bm{A}')$:
\begin{equation}
    Q_{\xi}(\bm{x}, \bm{A}) = r + \gamma \mathbb{E}[Q_{\xi}(\bm{x}', \bm{A}')], ~~\bm{A}'  \sim \pi_\phi(\bm{A}'|\bm{x}')
\end{equation}
where $\gamma$ is the discount factor for future rewards.
For a large state space, the value can be estimated with a differential function approximator $Q_{\xi}(\bm{x}, \bm{A})$, with parameters $\xi$. 
\begin{figure}[!htb]
	\centering
	\includegraphics[width=1.0\columnwidth]{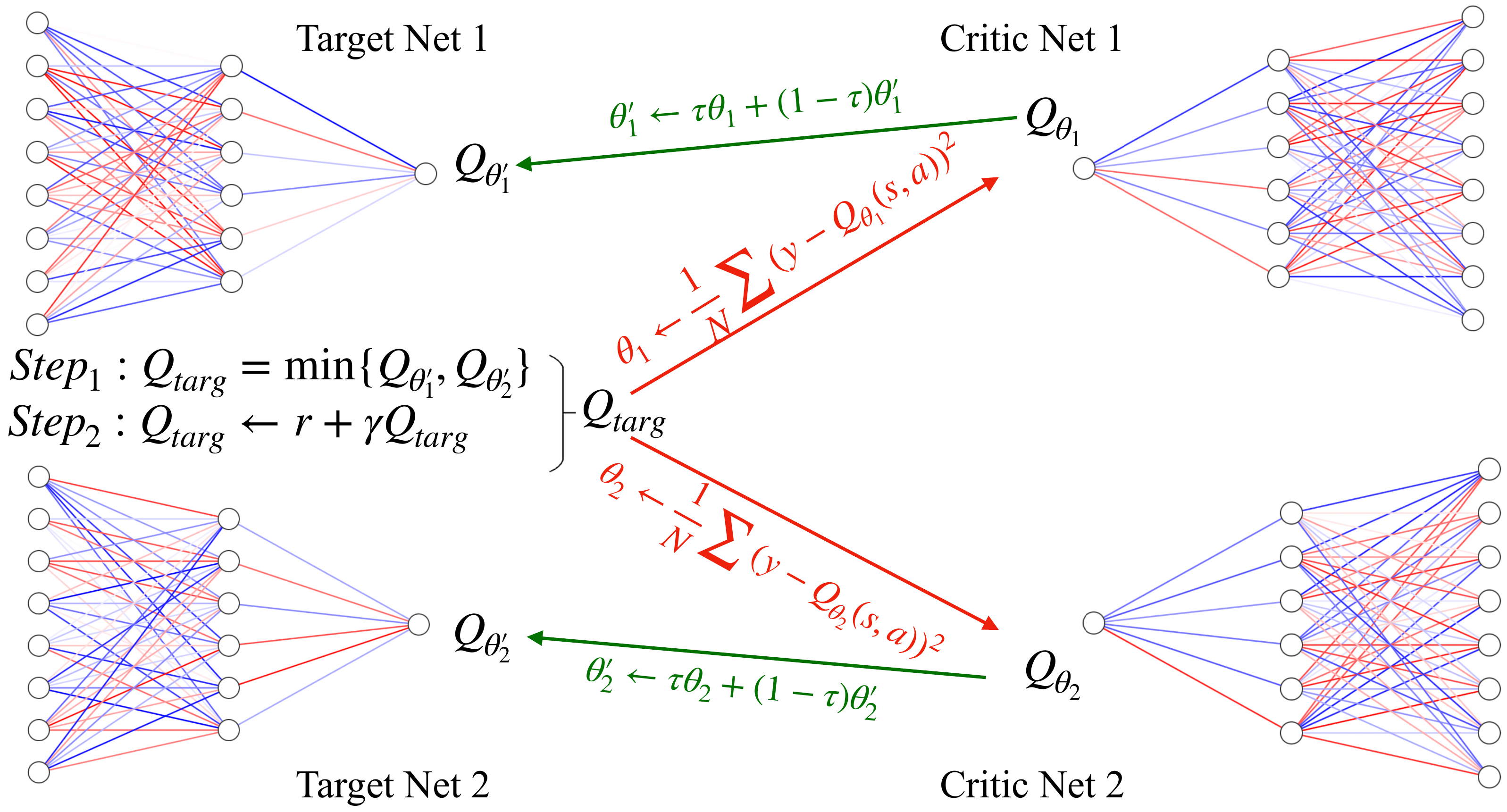}
	\caption{The critic design.}
	\label{fig:critic_net}
		\vspace{-0.3cm}
\end{figure}
In deep Q-learning, the network is updated by using temporal difference learning with a critic network $Q_{\xi}(\bm{x}, \bm{A})$ to maintain a fixed objective $y$ over multiple updates:
\begin{equation}
    y = r + \gamma  Q_{\xi}(\bm{x}, \bm{A}), ~~\bm{A}  \sim \pi_{\phi}(\bm{A}|\bm{x}),\label{targetnet}
\end{equation}
where the actions are selected from a target actor network $\pi_{\phi}$.  
\paragraph{Target Networks}
A well-known tool to achieve stability in deep reinforcement learning and reducing the function approximation error is introducing {\it target networks}  \cite{van2016deep}. In Fig. \ref{fig:critic_net}, we consider two target networks, i.e., $Q_{\xi'_1}$ and $Q_{\xi'_2}$, and two critic networks, i.e., $Q_{\xi_1}$ and $Q_{\xi_2}$.
The Clipped Double DQN \cite{van2016deep} uses the target networks (i.e. $Q_{\xi'_1}$ and $Q_{\xi'_2}$) by taking the minimun between  two value estimates:
\begin{equation}
\begin{aligned}\label{targetq}
y = r + \gamma  \min_{i=1,2}Q_{\xi'_i}(\bm{x}, \bm{A}),  
\end{aligned}
\end{equation}
where the value target $y$ cannot introduce any additional overestimation compared to the standard Q-learning target. This improves the estimation for the critic networks.
\paragraph{Critic Networks}
After receiving the target Q values, the critic networks update their own networks' parameters by: 
\begin{equation}
\begin{aligned}\label{critic_net}
&\xi_1 \leftarrow \arg\min_{\xi_1} \frac{1}{N}\sum(y - Q_{\xi_1}(\bm{x}, \bm{A} ) )^2,\\
&\xi_2 \leftarrow \arg\min_{\xi_2} \frac{1}{N}\sum(y - Q_{\xi_2}(\bm{x}, \bm{A} ) )^2
\end{aligned}
\end{equation}
where $N$ is the batch size, and $y$ is given by the target networks in \eqref{targetq}. After the critic networks are updated,  
the weights of a target networks are updated by some proportion $\tau$ by the critic networks at each time step:
\begin{equation}
\begin{aligned}\label{target1}
   \xi'_1 \leftarrow \tau \xi_1 + (1-\tau) \xi'_1,\\
   \xi'_2 \leftarrow \tau \xi_2 + (1-\tau) \xi'_2.
\end{aligned}
\end{equation}
where $\xi_1$ and $ \xi_2$ are the parameters of critic networks corresponding to Eq.\eqref{critic_net} and $\xi'_1$ and $\xi'_2$ are the parameters of target networks corresponding to Eq.\eqref{targetnet} 
Here, the target networks and critic networks are alternatively updated by each other.

\subsubsection{Constrained Actor Design}
After the critic is well defined, we can further define the actor network, and then introduce the constrained action space for the actor network. Typically, we train the action network to maximize the critic network, i.e.
\begin{equation}
\begin{aligned}
\phi \leftarrow \arg\max_{ {\phi}} Q_{\xi_1}(\bm{x}_{t-1} , \pi_{\phi}( \bm{x}_{t-1})).
\end{aligned}
\end{equation}
where $\phi$ is the parameters of  the action network.
Here, we can use either  $Q_{\xi_1}$ (or $Q_{\xi_2}$)  to guide $\pi_{\phi}(\cdot)$ to update ${\phi}$.
We call an action $\bm{A}_{t} \sim \pi_{\phi}( \bm{x}_{t-1})$ a feasible solution if $\bm{A}_t$ satisfies all its constraints, $\chi_t$. Thus, the policy ${\pi}_\phi$ is obtained by maximizing the critic network and satisfying $\chi_t$:
\begin{equation}
\begin{aligned}\label{constrqlearningpr}
\max_{ {\phi}} Q_{\xi_1}(\bm{x}_{t-1}, \pi_{\phi}(\bm{x}_{t-1}))
~~~s.t.~ \bm{A}_t \in \chi_t. 
\end{aligned}
\end{equation}
where $\bm{A}_t$ is taken according to policy $ \bm{A}_t = \pi_{\phi}(\bm{x}_{t-1})$.

\subsection{Primal-Dual Constrained RL Framework}
The actor network $\pi_{\phi}(\cdot)$ involves $  {\bm{g}}^p_t, {\bm{g}}^q_t, {\bm{p}} _{ch,t}, {\bm{p}}_{dis,t} $, which are linearly constrained in \eqref{DDCOPF_cst1} - \eqref{DDCOPF_cst7}. To write the linear equality constraints in a compact way, we consider 
\begin{equation}
\begin{aligned}
\bm{L} \pi_{\phi} (\bm{x}) = \bm{b}  
\end{aligned}
\end{equation}
Likewise, the linear inequality constraints are given by 
\begin{equation}
\begin{aligned}
\bm{K} \pi_{\phi} (\bm{x}) \preceq \bm{c}  
\end{aligned}
\end{equation}
Or equivalently,
\begin{equation}
\begin{aligned}
[\bm{K} \pi_{\phi} (\bm{x})  -  \bm{c}]_+   = \bm{0}
\end{aligned}
\end{equation}
where $[a]_+ = \max\{a, 0\}$.
We can summarize the optimization problem of the constrained reinforcement learning as
\begin{equation}
\begin{aligned}\label{qorg}
\min_{\phi} ~~&  -Q_{\xi_1}(\bm{x}', \pi_{\phi}(\bm{x})) \\
s.t.~~& \bm{L} \pi_{\phi} (\bm{x}) - \bm{b} =  \bm{0} \\
~~&  \bm{K} \pi_{\phi} (\bm{x}) - \bm{c}  \preceq \bm{0}
\end{aligned}
\end{equation}
where $Q_{\xi_1}$ is a non-convex, non-differentiable and Lipschitz continuous function. Both $  \bm{L} \pi_{\phi} (\bm{x}) - \bm{b} =  \bm{0},  \bm{K} \pi_{\phi} (\bm{x}) - \bm{c}  \preceq \bm{0}$ are linear, but $\pi_\phi(\bm{x})$ is non-convex, non-differentiable and Lipschitz continuous function. 
\begin{enumerate}
    \item Primal Problem by SGD for $T$ loops:
 \begin{equation}
\begin{aligned}\label{qlag}
 & \min_{\phi}\mathcal{L}^\alpha_\phi =      \min_{\phi} - Q_{\xi_1}(\bm{x}', \pi_{\phi}(\bm{x})) +   \bm{\lambda}^{\top}   [  \bm{L} \pi_{\phi} (\bm{x}) - \bm{b}]  \\
& + \bm{\mu}^{\top}    [\bm{K} \pi_{\phi} (\bm{x})  -  \bm{c}]_+ + \frac{{\alpha}^{\top}_I}{2}  \norm {\bm{L} \pi_{\phi} (\bm{x}) - \bm{b}  }_2^2 \\
& +\frac{{\alpha}^{\top}_F}{2}  \norm {[\bm{K} \pi_{\phi} (\bm{x})  -  \bm{c}]_+ }_2^2
\end{aligned}
\end{equation}  
where $\bm{x}'$ denotes the subsequent state.
\item Dual Problem:
 \begin{equation}
\begin{aligned}\label{qdual}
 &   \bm{\lambda}^{k+1}  =    \bm{\lambda}^{k}  +   {\alpha}_\lambda     [\bm{L} \pi_{\phi} (\bm{x}) - \bm{b}]  \\
&  \bm{\mu}^{k+1}  =  \bm{\mu}^{k}  +    {\alpha}_\mu   [\bm{K} \pi_{\phi} (\bm{x})  -  \bm{c}]_+
\end{aligned}
\end{equation} 
\end{enumerate}

\section{Cplx-GCN-based Actor Networks}
A reinforcement learning algorithm continuously interacts with the environment, which provides the time-series of the system states. The physics of the grid implies that the grid state variables correlation is a function of the grid topological and electrical characteristics.  It has been amply documented at this point in time that the best way to leverage the knowledge of the underlying grid is to use graph convoltional neural networks.

In this work, we consider a graph signal $\bm{x} \in \mathbb{R}^{\abs{\mathcal{V}}}$, where each entry $\left[\bm{x}\right]_i$ represents the voltage phasor at bus $i \in \mathcal{V}$. The set $\mathcal{N}_i$ denotes the nodes connected to node $i$. The graph shift operator (GSO) $\mathbf{S} \in \mathbb{R}^{\abs{\mathcal{V}} \times \abs{\mathcal{V}}}$ linearly combines values of the graph signal's neighbors. Operations such as filtering, transformation, and prediction are closely related to the GSO. In this work, we focus on complex symmetric GSOs, meaning $\mathbf{S} = \mathbf{S}^\top$. This is relevant for our power grid application, where $\mathbf{S} = \mathbf{Y}$ \cite{ramakrishna2021grid}. 

\begin{figure}[!htp]
  \center
    \includegraphics[width=0.42\textwidth]{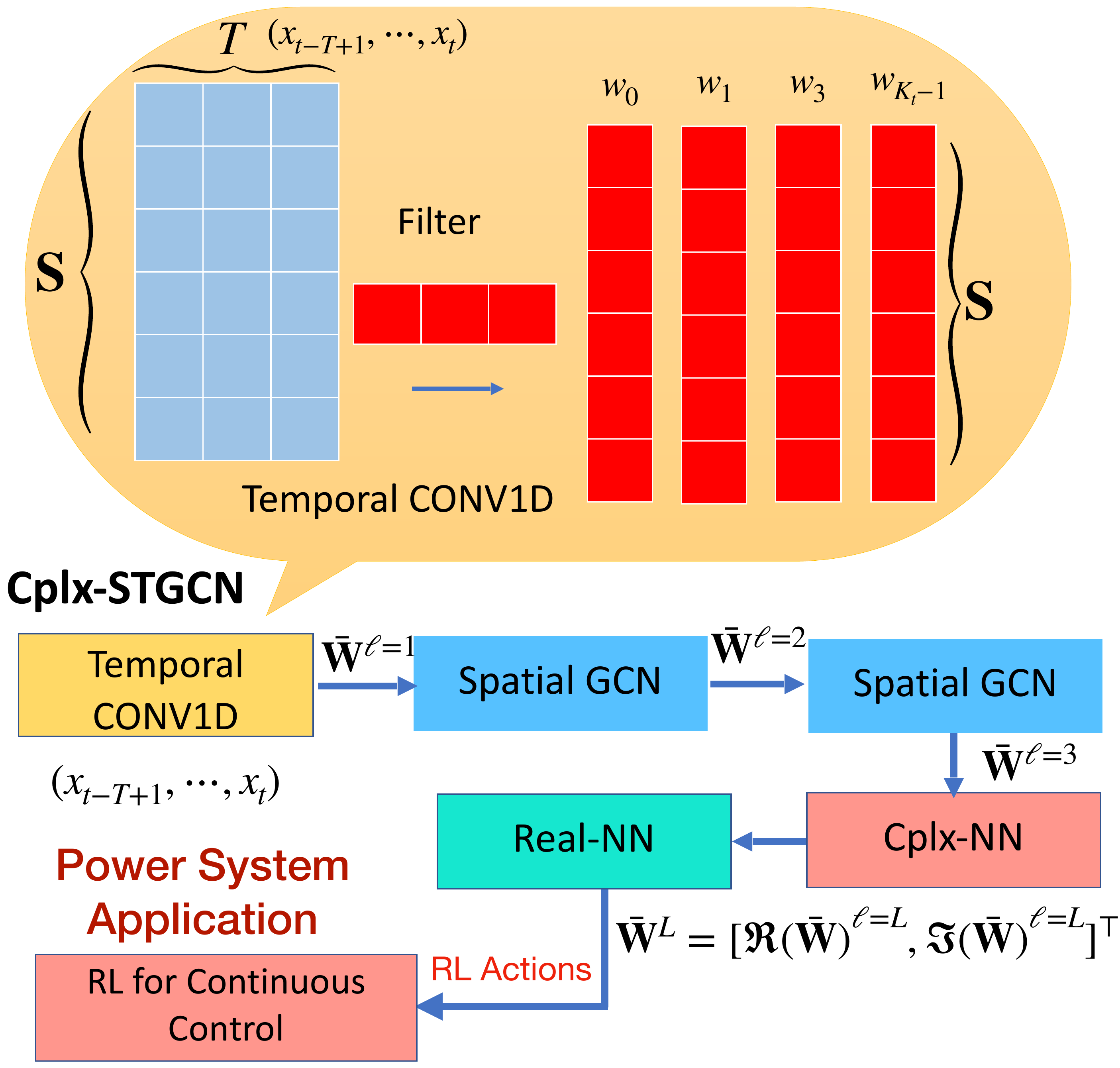}
  \caption{The Architecture of Cplx-STGCN.}\label{architecture_cgcn}
  \vspace{-0.4cm}
\end{figure}
As shown in Fig. \ref{architecture_cgcn}, the temporal convolutional layer contains a 1-D CNN with a width-$T$ kernel with $K_t$ output channels. In this work, we consider the input state $\bm{x}_t \in \mathbb{C}^{\abs{\mathcal{V}} \times 2}$ with two channels, i.e.,  $\bm{v}_t$ and $\bm{soc}_t$.
The convolutional kernel $\mathbf{\Gamma}\in \mathbb{C}^{2T\times K_t}$ is designed to map the input $\mathbf{X}\in \mathbb{C}^{\abs{\mathcal{V}}\times 2T}$ into a output graph signal with $C_t$ channels $\bar{\mathbf{X}}\in \mathbb{C}^{\abs{\mathcal{V}}\times K_t}$. Therefore, we define the temporal convolution as,
\begin{equation}
 \bar{ \mathbf{X}} =  \mathbf{\Gamma}*_{\mathcal{T}} \mathbf{X},
\end{equation}
where each column of  $ [\bar{ \mathbf{X}}]_\tau$ is defined as $ \bar{ \bm{x}}_\tau, \tau = 0, 1, \cdots, K_t-1$.
After the temporal convolutional layer, we are ready to put $ \bar{ \mathbf{X}}$ into the spatial layer.
Based on  \cite{wu2022complex}, we can design the following transfer functions and neuron:
\begin{equation}
\begin{split}\label{stkernal2}
\mathbb{H}(\mathbf{S}, z) = \sum_{t=0}^{K_t-1} \sum_{k=0}^{K-1} h_{k, t} \mathbf{S}^k z^{-t},\\
\bar{\bm{w}}_t = \sigma[\bm{w}_t] =  \sigma\left[\sum_{k=0}^{K-1} \sum_{\tau=0}^{K_t-1} h_{k, \tau} \mathbf{S}^k \bm{x}_{t-\tau}\right].
\end{split}
\end{equation}
Accordingly, the graph signal $\bar{\bm{w}}_t$ from the spatial feature extraction layer (see Fig. \ref{architecture_cgcn}) is:
\begin{equation}
\begin{split}\label{graph_signal}
\bar{\bm w_t} =  \operatorname{CReLU} \left[\sum_{k=0}^{K-1} \sum_{\tau=0}^{K_t-1} h_{k, \tau} \mathbf{S}^k \bar{\bm{x}}_{t-\tau}\right],
\end{split}
\end{equation}
By combining the temporal and spatial convolutions at each layer,  the multiple output channels of the Cplx-STGCN layer ($\ell = 1$) are expressed as
\begin{equation}
\begin{split}\label{graph_signal}
\bar{\mathbf{W}}_{t,\ell=1}  = \operatorname{CReLU}(\mathbf{H}*_{\mathcal{G}} (\mathbf{\Gamma}*_{\mathcal{T}} \mathbf{X}_t)),
\end{split}
\end{equation}
where $\mathbf{H}$ and $\mathbf{\Gamma}$  are the trainable parameters. We  denote \eqref{graph_signal} by the feature extraction  layer.

The graph neural network perception  is:
\begin{equation}
\bar{\bm{w}}=\sigma[\bm{w}] = \sigma\left[\sum_{k=0}^{K-1} h_{ k} \mathbf{S}^{k} \bm{x}\right]
\end{equation}
where $\bm{x}\in \mathbb{C}^{\abs{\mathcal{V}}}$, $h_k\in \mathbb{C}$, $\mathbf{S}^k\in \mathbb{C}^{\abs{\mathcal{V}}\times \abs{\mathcal{V}}}$ and $\bm{w}\in \mathbb{C}^{\abs{\mathcal{V}}}$ are the complex values. Since  $\sigma(\cdot)$ takes as input complex values, there is significant flexibility in defining this operator in  the complex plane. In the following, we refer to  Complex ReLU (namely CReLU) as the simple complex activation that applies separate ReLUs on both of the real and the imaginary part of a neuron, i.e: 
\begin{equation}
\operatorname{CReLU}(\bm{w})=\operatorname{ReLU}(\Re(\bm{w}))+\mathfrak{j} \operatorname{ReLU}(\Im(\bm{w})).
\end{equation}

Spatio-Temporal GCN are a special case of {\it multiple features GCN}. Specifically, let $\mathbf{X} = [\bm{x}^1, \cdots, \bm{x}^F]$ and let us refer to the multiple channel outputs as $\mathbf{W} = [\bm{w}^1, \cdots, \bm{w}^G]$, where $F$ is the number of input features and $G$ is the number of output channels. A layer of  multiple features GCN operates as follows:
\begin{equation}\label{eq:MF-GCN}
\bar{\mathbf{W}} = \sigma[\mathbf{W}] =\sigma\left[\sum_{k=0}^{K-1} \mathbf{S}^{k} \times \mathbf{X} \times \mathbf{H}_{k}\right] = \operatorname{CReLU}(\mathbf{H}*_{\mathcal{G}} \mathbf{X}),
\end{equation}
where these matrices include $ G \times F$ coefficient matrix $\mathbf{H}_{k}$ with entries $\left[\mathbf{H}_{k}\right]_{{fg}}=h_{k}^{f g}$, and  $\mathbf{H}*_{\mathcal{G}}$ defines   the notion of graph convolution operator based on the concept of spectral graph convolution. 

\section{Convergence Analysis}
\subsection{Convergence of Value Functions}
We first focus on the convergence of the value function:
\begin{theorem}\label{thm1}
We make the following assumptions:
\begin{enumerate}
    \item Each state action pair is sampled an infinite number of times.
    \item The Markov decision process is finite.
    \item $\gamma \in [0, 1)$.
    \item $Q$ values are stored in a lookup table.
    \item $Q_{\xi_1}$ and $Q_{\xi_2}$ receive an infinite number of updates.
    \item The learning rates satisfy $\eta_t\in [0, 1], \sum_t \eta_t = \infty, \sum_t (\eta_t)^2 < \infty$ with probability 1 and $\forall (s, a) \neq (s_t, a_t), \eta_t = 0$.
    \item $\forall r, \mathbb{V}[r] < \infty$
\end{enumerate}
Then constrained TD3 will converge to the optimal value function $Q^*$, as defined by the Bellman optimality equation, with probability 1.
\end{theorem}
The primal-and-dual update only applies to the actor function instead of the policy function. Therefore, the proof is shown in Section A of Supplementary Material \cite{fujimoto2018addressing} applied to the value function of the proposed CRL.  In Theorem  \ref{thm1}, $Q_{\xi_1}(\bm{x}_t, \bm{a}_t)$ converges to $Q^*(\bm{x}_t, \bm{a}_t)$. 

\subsection{Convergence of Actor Functions}

In the following, we analyze the convergence of the actor network after the critic network converges.
The policy function $\pi_\phi$ can be expressed by 
\begin{equation}
\begin{aligned}
\pi_\phi = \left\{
\begin{matrix}
&\bar{\mathbf{W}}_{\ell=1}  = \operatorname{CReLU}(\mathbf{H}*_{\mathcal{G}} (\mathbf{\Gamma}*_{\mathcal{T}} \mathbf{X}))\\
&\bar{\mathbf{W}}_{\ell+1}=  {\operatorname{CReLU}\left(\Theta_\ell^{cplx} * {\bar{\mathbf{W}}_{\ell}}\right)}, ~~ 1\le\ell \le L-1\\
&\bm{a} = \text{tanh}\left(\Theta_{L}^{cplx} *
\bar{\mathbf{W}}_{L}
\right)
\end{matrix}
\right.
\end{aligned}
\end{equation}
It is obvious that $\pi_\phi(\bm{x})$ is non-convex, non-differentiable due to the activation function, but subderivative. Recall that we have the primal-and-dual method for the actor network:


\begin{assumption}
The primal update for \eqref{qlag} can always find the local optimal solution $\phi^*$ for a long $T$ loop:
 \begin{equation}
\begin{aligned}\label{qlag}
 &  \phi^* =      \argmin_{\phi} - Q^*(\bm{x}', \pi_{\phi}(\bm{x})) +   \bm{\lambda}^{\top}   [\bm{L} \pi_{\phi} (\bm{x}) - \bm{b}]  \\
& + \bm{\mu}^{\top}    [\bm{K} \pi_{\phi} (\bm{x})  -  \bm{c}]_+ + \frac{{\alpha}^{\top}_\lambda}{2}  \norm {  [\bm{L} \pi_{\phi} (\bm{x}) - \bm{b}]}_2^2 \\
& +\frac{{\alpha}^{\top}_\mu}{2}  \norm {[\bm{K} \pi_{\phi} (\bm{x})  -  \bm{c}]_+ }_2^2
\end{aligned}
\end{equation}  
\end{assumption}
This assumption was already proved by in \cite{zou2019sufficient}.

\begin{assumption}
We define the unaugmented Lagrangian  $ \mathcal{L}_\phi$ has a saddle point $(\phi^*, \bm{\lambda}^*, \bm{\mu}^*)$.
\end{assumption}

\begin{definition}
We define the equality residual $\bm{r}_\lambda = \bm{L} \pi_{\phi} (\bm{x}) - \bm{b} $ and the inequality residual $\bm{r}_\mu = [\bm{K} \pi_{\phi} (\bm{x}) - \bm{c}]_+$.
\end{definition}

\begin{lemma}
Let $(\phi^*, \bm{\lambda}^*, \bm{\mu}^*)$ be a saddle point for $\mathcal{L}^\alpha_\phi$, and define 
\begin{equation}
\begin{aligned}
V^k = \frac{1}{\alpha_\lambda}\norm{\bm{\lambda}^k - \bm{\lambda}^*}_2^2  + \frac{1}{\alpha_\mu}\norm{\bm{\mu}^k - \bm{\mu}^*}_2^2  
\end{aligned}
\end{equation} 
$V^k$ decreases until $\bm{r}_\lambda \rightarrow 0$ and $\bm{r}_\mu \rightarrow 0$.
\end{lemma}
With Assumption III.1 and Lemma 1, we can conclude that both primal and dual updates converge into a saddle point.

\section{Case Study: Primal-Dual CRL Implementation for SDOPF}
We implement the proposed algorithm  (\eqref{qlag} and \eqref{qdual}) for the multi-stage stochastic dynamic optimal power flow problem \eqref{DDCOPF_all}. Then, we summarize the detailed steps of the primal and dual updates in Algorithm \ref{algVPF}.
\subsection{Primal-Dual  SDOPF Formulation}
With the above primal-dual framework , we aim to   train the constrained  policy function $\pi_\phi(\cdot)$ for SDOPF. In particular, we  define power generations $ {\bm{g}}^p_t, {\bm{g}}^q_t$, the BESS charging power $\bm{p}_{ch,t}$ and discharging power $\bm{p}_{dis,t}$ by
the actions $\bm{A}_t = [\hat{\bm{a}}_t, \cdots, \hat{\bm{a}}_{t+T-1}]$ as
\begin{equation}\label{mappingchdis}
\begin{aligned}
&\hat{\bm{a}}_t \triangleq  [\pi_{\phi}(\bm{x}_{t-1})]_{t}, ~~ \hat{\bm{a}}_t \triangleq [\hat{\bm{g}}_t^p;\hat{\bm{g}}_t^q; \hat{\bm{p}}_{ch,t}; \hat{\bm{p}}_{dis,t}]^\top,    \\
&  \hat{\bm{g}}^p_t \triangleq \hat{\bm{a}}_{t, 1},  ~~ \hat{\bm{g}}^q_t \triangleq \hat{\bm{a}}_{t, 2}, ~~ \hat{\bm{p}}_{ch,t} \triangleq   \hat{\bm{a}}_{t, 3},~~ \hat{\bm{p}}_{dis,t} \triangleq  \hat{\bm{a}}_{t, 4}, \\
& \bm{g}^p_t \triangleq (1-\hat{\bm{g}}^p_t) \underline{\bm{g}}^p+   \hat{\bm{g}}^p_t \overline{\bm{g}}^p, ~~ \bm{g}^q_t \triangleq (1-\hat{\bm{g}}^q_t) \underline{\bm{g}}^q+   \hat{\bm{g}}^q_t \overline{\bm{g}}^q, \\
& {\bm{p}}_{ch, t} \triangleq \hat{\bm{p}}_{ch,t} \bm{P}^{ch}_{rated},~~  {\bm{p}}_{dis, t} \triangleq \hat{\bm{p}}_{dis,t} \bm{P}^{dis}_{rated}.     
\end{aligned}
\end{equation}
We also have the predicted voltage magnitudes $\abs{\bm{v}}$ as:
\begin{equation}\label{predV}
\begin{aligned}
\abs{\hat{\bm{v}}_t} = [P_{\omega}(\bm{x}_{t-1})]_{t},  \abs{\bm{v}_t^p} = (1-\abs{\hat{\bm{v}}_t })\underline{\bm{v}} + \abs{\hat{\bm{v}}_t} \overline{\bm{v}},
\end{aligned}
\end{equation}
where $\abs{\bm{v}_t^p}$ is constrained to be equal to the ground-truth one, i.e., $\abs{\bm{v}_t}$. Therefore, we replace $\abs{\bm{v}_t}$ in \eqref{DDCOPF_cst1} and \eqref{DDCOPF_cst2} with $\abs{\bm{v}_t^p}$, with an additional constraint:
\begin{equation}\label{predV}
\begin{aligned}
\abs{\bm{v}_t^p} - \abs{\bm{v}_t} = 0.
\end{aligned}
\end{equation}
With the definition of \eqref{mappingchdis}, the augmented Lagrangian is:
\begin{equation}\small
\begin{aligned}
&\min_{ {\phi}} \mathcal{L}_{\phi} = - Q_{x'_i}(\bm{x}_t, \pi_{\phi}(\bm{x}_{t-1})) + \sum_{t = \tau+1}^{\tau+T} \bigg( \bm{\lambda}_{1,t}^\top \big( \mathbf{M}_{b}\bm{p}_{dis,t}-   \\
&  \mathbf{M}_{b} \bm{p}_{ch,t}  + \mathbf{M}_{g}\bm{g}^p_{t} -   \bm{d}^p_t - \Re\{D(\bm{v}^p_t (\bm{v}_t^p)^H \mathbf{Y}^H)\} \big) +  \bm{\lambda}_{2,t}^\top \big(\mathbf{M}_{g}\bm{g}^q_{t}   \\
& -   \bm{d}^q_t - \Im\{D(\bm{v}^p_t (\bm{v}_t^p)^H \mathbf{Y}^H)\} \big) + \bm{\lambda}_{3,t}^\top \big(\bm{soc}_{t} -   \bm{soc}_{t-1} +    \frac{\Delta t}{E_{cap}}\\
&   \big(    \eta_{ch} \bm{p}_{ch,t}  -  \frac{\bm{p}_{dis,t}}{ \eta_{dis}}  \big)\big) + \bm{\mu}_{1,t}^\top (  \bm{soc}_{t-1}  - \frac{\Delta t \big(    \eta_{ch} \bm{p}_{ch,t} - \frac{\bm{p}_{dis,t} }{ \eta_{dis}}  \big)}{E_{cap}}    \\
&    - \bm{soc}_{max}) +\bm{\mu}_{2,t}^\top(  \bm{soc}_{min}  + \frac{\Delta t  \big(    \eta_{ch}  \bm{p}_{ch,t}  - \frac{\bm{p}_{dis,t} }{ \eta_{dis}} )}{E_{cap}}  -  \bm{soc}_{t-1}  )  \\
& + \bm{\lambda}^\top_{4,t}  (\abs{\bm{v}_t^p} - \abs{\bm{v}_t} )  +  \bm{\mu}^\top_{3,t}\Big( \abs{\bm{v}_t^p} - \overline{\bm{v}}\Big) +  \bm{\mu}^\top_{4,t} \Big(\underline{\bm{v}} -    \abs{\bm{v}_t^p}  \Big)  \\
& +  \frac{\alpha_{1,t}}{2}  \Big\|\mathbf{M}_{b}(\bm{p}_{dis,t}-   \bm{p}_{ch,t}) +    \mathbf{M}_{g}  \bm{g}^p_{t}  -  \Im\{D(\bm{v}_t \bm{v}_t^H \mathbf{Y}^H)\} \Big\|      \\
&   + \frac{\alpha_{2,t}}{2}  \Big\|  \mathbf{M}_{g}  \bm{g}^q_{t} - \Im\{D(\bm{v}_t \bm{v}_t^H \mathbf{Y}^H)\} -   \bm{d}^q_t  \Big\|_2^2   \\
& +  \frac{\alpha_{3,t}}{2}    \Big\|\bm{soc}_t -     \bm{soc}_{t-1}  + \frac{\Delta t}{E_{cap}} \big(    \eta_{ch} \bm{p}_{ch,t}  - \frac{\bm{p}_{dis,t} }{ \eta_{dis}}  \big)\Big\|_2^2  \\
& +\frac{\alpha_{4,t}}{2} \norm{      \bm{soc}_{t-1}  - \frac{\Delta t}{E_{cap}} \big(    \eta_{ch} \bm{p}_{ch,t}  - \frac{\bm{p}_{dis,t} }{ \eta_{dis}}  \big)    - \bm{soc}_{max}}_2^2     \\
& + \frac{\alpha_{5, t}}{2} \norm{ \bm{soc}_{min} -   \bm{soc}_{t-1}  + \frac{\Delta t}{E_{cap}} \big(    \eta_{ch} \bm{p}_{ch,t}  - \frac{\bm{p}_{dis,t} }{ \eta_{dis}}  \big)    }_2^2\\
&  +\frac{\alpha_{6,t}}{2}\|  \abs{\bm{v}_t^p}  - \abs{\bm{v}_t}\|_2^2  +\frac{\alpha_{7,t}}{2}\|   \abs{\bm{v}_t^p} - \overline{\bm{v}}\|_2^2 +\frac{\alpha_{8,t}}{2}\|    \abs{\bm{v}_t^p} - \underline{\bm{v}} \|_2^2 
\bigg) \label{constrqlearning}
\end{aligned}
\end{equation}
where $\bm{\lambda}_{1,t}$, $\bm{\lambda}_{2,t}$, $\bm{\lambda}_{3,t}$, $\bm{\lambda}_{4,t}$, $\bm{\mu}_{1,t}$, $\bm{\mu}_{2,t}$, $\bm{\mu}_{3,t}$ and $\bm{\mu}_{4,t}$ are the dual variable vectors, and $\alpha_{1,t}$, $\alpha_{2,t}$, $\alpha_{3,t}$, $\alpha_{4,t}$, $\alpha_{5,t}$ $\alpha_{6,t}$, $\alpha_{7,t}$  and  $\alpha_{8,t}$ are positive scalars that penalize the augmented terms. 
The above problem is different to the optimization problem in \eqref{DDCOPF_all} due to the following facts:
\begin{enumerate}
    \item In \eqref{constrqlearning}, we replace $\bm{g}^p$, $\bm{g}^q$, $\bm{p}_{dis}$ and $\bm{p}_{ch}$  together with the bound constraints, $ \underline{\bm{g}}^p \le \bm{g}^p_t \le \overline{\bm{g}}^p$, $ \underline{\bm{g}}^q \le \bm{g}^q_t \le \overline{\bm{g}}^q$,  $0\le \bm{p}_{ch,t} \le \bm{P}^{ch}_{rated}$,  $	0\le \bm{p}_{dis, t} \le  \bm{P}^{dis}_{rated}$ by  \eqref{mappingchdis}. Note that   $\pi_{\phi}(\bm{x}_{t-1}) $ is the output of the neural network by applying a sigmoid to the final layer, and thus $\pi_{\phi}(\bm{x}_{t-1})  \in [0, 1]$.  
    \item  $\bm{soc}$ are state variables in \eqref{DDCOPF_all} that need to be solved. However, in \eqref{constrqlearning}, $\bm{soc}_i$ are   the given training samples, which are utilized to bound the actions $\pi_{\phi}(\bm{x}_{t-1}) $.
    \item For the voltage magnitude bound, i.e., $\abs{\bm{v}}$, we need to predict it by  a independent GCN given the states $\bm{x}_t$, i.e.,  $\abs{\bm{v}^p}  $. We also need to make a constraint for the predicted voltage magnitude equal to the ground-truth one, i.e., $\abs{\bm{v}^p}  = \abs{\bm{v}} $.
\end{enumerate}

The primal dual update requires minimizing the Lagrangian function and then maximizing the dual function  With the definition of \eqref{mappingchdis}, the dual variables gradient update is: 
\begin{equation}\small
\begin{aligned} 
&\bm{\lambda}_{1,t}^{k+1}   \leftarrow \bm{\lambda}_{1, t}^{k}  + \alpha_{1,t}(\mathbf{M}_{b}\bm{p}_{dis,t} - \mathbf{M}_{b}\bm{p}_{ch,t} + \mathbf{M}_{g}\bm{g}^p_t   \\
&- \bm{d}_t -  \Re\{D(\bm{v}_t \bm{v}_t^H \mathbf{Y}^H)\} ), \forall t \in [\tau, \tau+T]  \\
&\bm{\lambda}_{2,t}^{k+1}   \leftarrow \bm{\lambda}_{2, t}^{k}  + \alpha_{2,t}(  \mathbf{M}_{g}^q\bm{g}_t    - \bm{d}_t -  \Im\{D(\bm{v}_t \bm{v}_t^H \mathbf{Y}^H)\}), \\
&\forall t \in [\tau, \tau+T]  \\
 &  \bm{\lambda}_{3, t}^{k+1}    \leftarrow \bm{\lambda}_{3, t}^{k}  + \alpha_{3, t} \bigg( \bm{soc}_{t}  - \bm{soc}_{t-1}       +    \frac{\Delta t}{E_{cap}} \Big(    \eta_{ch}  \bm{p}_{ch,t}\\
& - \frac{\bm{p}_{dis,t}}{ \eta_{dis}}  \Big)\bigg),  \forall t \in [\tau, \tau+T]  \notag
\end{aligned}
\end{equation}
\begin{equation}\small
\begin{aligned} 
&  \bm{\lambda}_{4, t}^{k+1}  \leftarrow  \bm{\lambda}_{4, t}^{k} + \alpha_{6, t}\Big( \abs{\bm{v}_t^p} - \abs{\bm{v}_t} \Big) , \forall t \in [\tau+1, \tau+T] \\
& \bm{\mu}_{1, t}^{k+1}  \leftarrow  \bm{\mu}_{1, t}^{k}  +  \alpha_{4, t} \Big[  \bm{soc}_{t-1}  - \frac{\Delta t}{E_{cap}} \big(    \eta_{ch} \bm{p}_{ch,t}  - \frac{\bm{p}_{dis,t} }{ \eta_{dis}}  \big)    \\
&-   \bm{soc}_{max}  \Big]_+   ,   \forall t \in [\tau+1, \tau+T] \\
& \bm{\mu}_{2, t}^{k+1}  \leftarrow  \bm{\mu}_{2,t}^{k} +  \alpha_{5, t} \Big[ \bm{soc}_{min}   -  ( \bm{soc}_{t-1}  - \frac{\Delta t}{E_{cap}} \big(    \eta_{ch} \bm{p}_{ch,t}\\
&- \frac{\bm{p}_{dis,t} }{ \eta_{dis}}  \big)    \Big]_+,  \forall t \in [\tau+1, \tau+T] \\
&  \bm{\mu}_{3, t}^{k+1}  \leftarrow \bm{\mu}_{3, t}^{k} + \alpha_{7, t} \Big[ \abs{\bm{v}_t^p}    - \overline{\bm{v}} \Big]_+, \forall t \in [\tau+1, \tau+T]\\
&  \bm{\mu}_{4, t}^{k+1}  \leftarrow    \bm{\mu}_{4, t}^{k} + \alpha_{8, t} \Big[ \underline{\bm{v}} - \abs{\bm{v}_t^p} \Big]_+, \forall t \in [\tau+1, \tau+T]
\label{dualupdate2}
\end{aligned}
\end{equation}
where  $\bm{\lambda}_{1,t}^{k+1} $,  $\bm{\lambda}_{2,t}^{k+1} $, $\bm{\lambda}_{3,t}^{k+1} $, $\bm{\lambda}_{4, t}^{k+1}$, $\bm{\mu}_{1, t}^{k+1}$, $\bm{\mu}_{2, t}^{k+1}$, $\bm{\mu}_{3, t}^{k+1}$ and $\bm{\mu}_{4, t}^{k+1}$ are updated by batch samples. We conduct the primal-dual update alternatively to optimize $\phi$ of $\pi_{\phi}(\cdot)$ while enforcing the feasibility of both equality and inequality constraints. 

\subsection{Constrained Reinforcement Learning Algorithm}

The training process of the constrained reinforcement learning algorithm, described above, is summarized in Algorithm  \ref{algVPF}. In steps 1-3, we initialize the parameters of double critic networks $Q_{\xi_1}$, $Q_{\xi_2}$, double target networks $Q_{\xi'_1}$, $Q_{\xi'_2}$, and the actor networks $\pi_{\theta}$. Steps 5-8 represent the process of data sampling and storing transition tuple  $  (\bm{x}_{t-1}, \bm{A}_{t}, r_{t}, \bm{x}_{t})$. In Steps 9-11, we update the 
critic networks, and we update the actor network in Steps 12-13 and the dual variables in Steps 15-16.  In Steps 19-25, we update the state of charge $\bm{soc}$ that should be projected within [0, 1]. In Step 26, we fix $\bm{g}^p_{t}, \bm{g}^q_{t}, \bm{p}_{ch, t}, \bm{p}_{dis, t}$ and solve power flow equations to obtain $\bm{v}_t$. Specifically, we fixed the power generations on the non-slack buses, i.e. ${g}_{i, t}^p$ and ${g}_{i, t}^q$ $\forall i\in \mathcal{G}_n$, and keep the voltage magnitude and angle on the slack bus  1 p.u. and 0, respectively. Then, we utilize the Newton's method to solve $\bm{v}_t$.
  \begin{algorithm} [!htb]\small
      \caption{Constrained Reinforcement Learning for multi-stage SDOPF}\label{algVPF}
   Initialize critic network $Q_{\xi_1}$, $Q_{\xi_2}$, and actor network $\pi_\phi$ with random parameters $\xi_1$, $\xi_2$ and $\phi$\;
   Initialize target networks $\xi'_1 \leftarrow \xi_1$, $\xi'_2 \leftarrow \xi_2$, $\phi'  \leftarrow \phi$\;
    Initialize replay buffer $\mathcal{B}$, the state $\bm{x}_0$ and set primal and dual update periods $pu$ and $du$\;
          \For{$t=1: T$}
        {
         \tcc{This the sampling processing}
         Select action  $\bm{A}_{t} \sim \pi_{\phi}(\bm{x}_{t-1})$\;
         Observe reward $r_{t}$  by Eq. \eqref{rwd1}\;
         Obtain the new state $\bm{x}_{t} = \textbf{env}(\bm{A}_{t})$ by taking action  $\bm{A}_{t}$\;
         Store  transition tuple  $  (\bm{x}_{t-1}, \bm{A}_{t}, r_{t}, \bm{x}_{t})$  in $\mathcal{B}$\;
            \tcc{This the training processing}
         Sample mini-batch of $N$ transitions $  \{(\bm{x}_{n-1}, \bm{A}_{n}, r_{n}, \bm{x}_{n})|n = 1,\cdots, N\}$ from $\mathcal{B}$\;
         $y \leftarrow r_t + \gamma  \min_{i=1,2}Q_{\xi'_i}(\bm{x}_{n}, \pi_{\phi}(\bm{x}_{n}))$\;
        
        Update critics:  $\xi_1 \leftarrow \arg\min_{\xi_1} \frac{1}{N}\sum(y - Q_{\xi_1}(\bm{x}_{n-1}, \bm{A}_{n}) )^2$ and $ \xi_2 \leftarrow \arg\min_{\xi_2} \frac{1}{N}\sum(y - Q_{\xi_2}(\bm{x}_{n-1}, \bm{A}_{n}) )^2$\;
        \If{t mod pu}
        {Update $\phi$ by the deterministic policy gradient: $\phi \leftarrow \phi  - \eta \nabla \mathcal{L}_\phi(\bm{x}_{n-1}, \bm{x}_{n})$, where $\eta$ is the learning rate and $\mathcal{L}_\phi$ is defined in \eqref{constrqlearning}\;
        Update target networks by \eqref{target1}\; }
        \If{t mod du}
        {
        Update the dual variables by \eqref{dualupdate2}.
        }
        }
      \Fn(){\textbf{env} ($\bm{A}_{t+1}$)}{
      Choose the first action, i.e., $\bm{a}_{t} =  [\bm{g}^p_{t}; \bm{g}^q_{t}; \bm{p}_{ch, t}; \bm{p}_{dis, t}]^\top $ from $  \bm{A}_{t} = [\bm{a}_{t}, \cdots, \bm{a}_{t'}]^\top, t'= t+T-1$ , and $[\bm{a}_{t+1}, \cdots, \bm{a}_{t+T-1}]^\top$ are scenario-based actions\;
       \For{$i\in \mathcal{B}$}{
      \If{$0<soc_{i, t}<1$}{
       ${soc}_{i,t+1}  \leftarrow {soc}_{i, t}     + \frac{\Delta t}{E_{cap}} \big(    \eta_{ch}  {p}_{ch,i, t}  - \frac{{p}_{dis,i,t} }{ \eta_{dis}}  \big)  $\;}
       \ElseIf{$soc_{i, t}=1$}{${p}_{ch,i, t}  \leftarrow 0$, ${soc}_{i,t+1}  \leftarrow {soc}_{i, t}    - \frac{\Delta t}{E_{cap}}      \frac{{p}_{dis,i,t} }{ \eta_{dis}}    $\;}
       \ElseIf{$soc_{i, t}=0$}{${p}_{dis,i, t}  \leftarrow 0$, ${soc}_{i,t+1}  \leftarrow {soc}_{i, t}     + \frac{\Delta t \eta_{ch}  {p}_{ch,i, t} }{E_{cap}}        $\;}
        }
     Obtain $\bm{v}_{t}$ by solving     Eqs.\eqref{DDCOPF_cst1} and \eqref{DDCOPF_cst2} with $\bm{a}_t$ fixed\;
     Return $\bm{x}_{t} = [\bm{v}_t, \bm{v}_{t-1}, \cdots,   \bm{v}_{t-T+1} ]^\top$\;
      }
\end{algorithm}

After training the GCN-policy function, we are ready to implement the GCN policy to forecast the control actions, i.e., $  \hat{\bm{g}}_t^p,\hat{\bm{g}}_t^q, \hat{\bm{p}}_{ch,t}, \hat{\bm{p}}_{dis,t} $, by feeding the previous state measurements $\bm{x}_{t-1}$. Then, the active and reactive power generations  $   {\bm{g}}_t^p, {\bm{g}}_t^q$, and the battery charging and discharging powers $ {\bm{p}}_{ch,t}, {\bm{p}}_{dis,t} $  are utilized to control the power systems in real time.

\section{Experimental Results}
\subsection{Experimental Setup}
Experiments were conducted on the IEEE 14-bus and 30-bus systems, each with two BESSs. The IEEE 14-bus system has BESSs located at Bus  9, while the IEEE 30-bus system has BESSs at Buses 13 and 22. These BESSs have a capacity $E_{cap}$ of 1000 and charge/discharge efficiencies ($\eta_{ch}$ and $\eta_{dis}$) of 0.98.
%
%
The time step is 18 seconds. The SOC is bounded in $[0, 1]$. For training of the constrained DRL we relied on PyTorch and used realistic demand profiles from the \href{https://www.ercot.com/gridinfo/load/load_hist}{\color{blue}Texas grid}.    We consider one wind power generation in IEEE 14-bus system, and three power generations in IEEE 30-bus system, where the real-world wind generations are collected from NREL Wind \cite{draxl2015wind}.

\begin{figure}[htbp]   %
\centering
\begin{minipage}{0.24\textwidth}
    \centering
    {\includegraphics[width=1\textwidth]{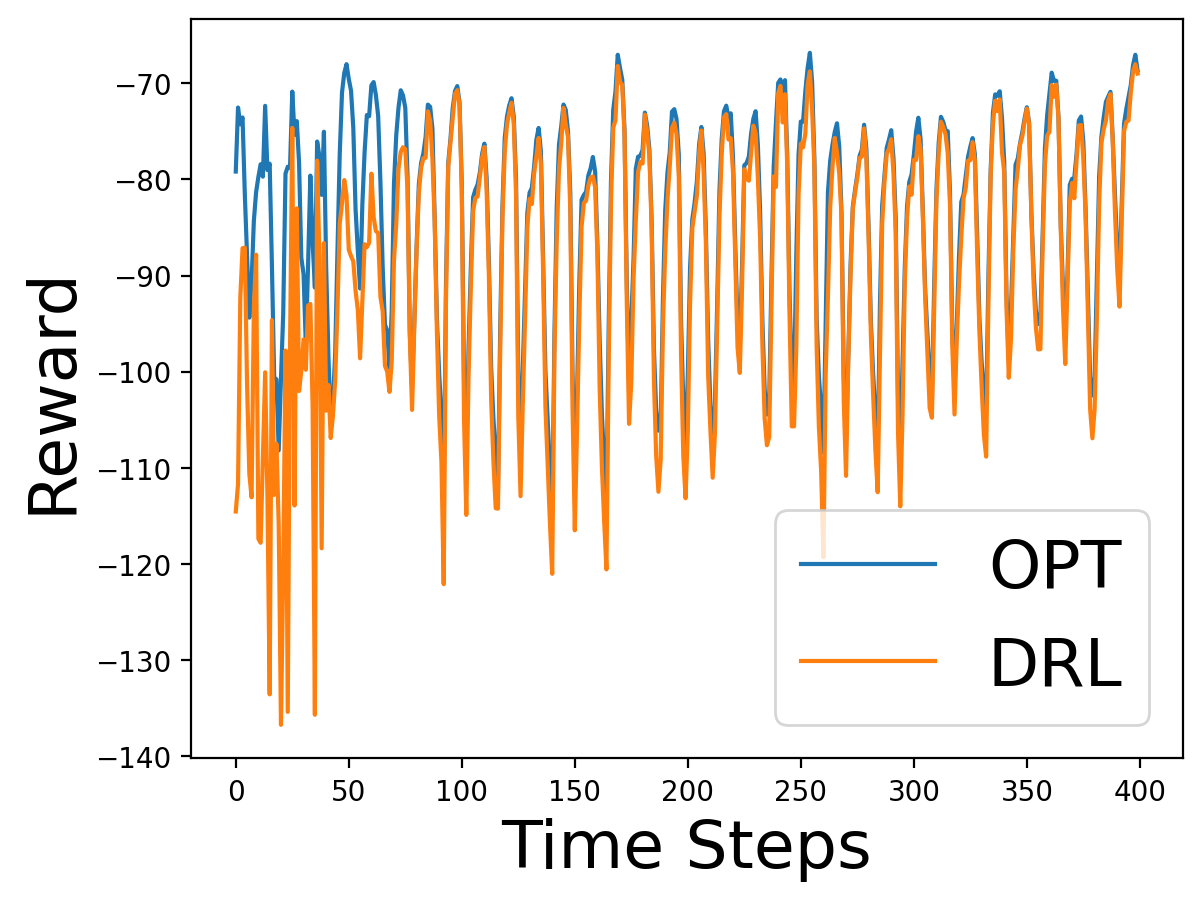}}
    \caption{The training curves of the  Constrained DRL and its corresponding optimal rewards by the optimization method in the IEEE 14-bus system.} \label{trainingcurve14bus}
\end{minipage}
\hspace{-0.1cm}
\begin{minipage}{0.24\textwidth}
    \centering
    {\includegraphics[width=1\textwidth]{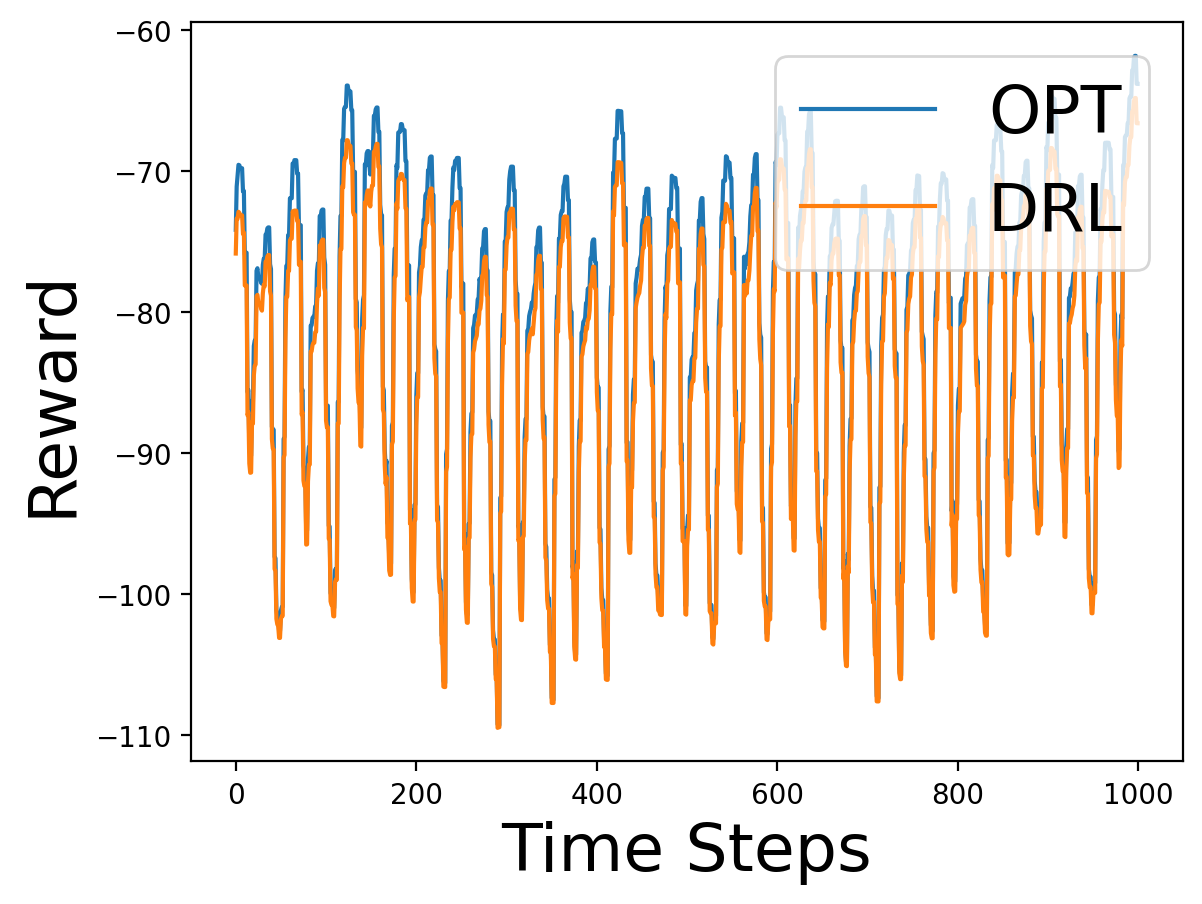}}
\caption{The testing curves of the  Constrained DRL and its corresponding optimal rewards by the optimization method in the IEEE 14-bus system.} \label{testingcurve14bus}
\end{minipage}
\hspace{-0.1cm}
\begin{minipage}{0.24\textwidth}
    \centering
    {\includegraphics[width=1\textwidth]{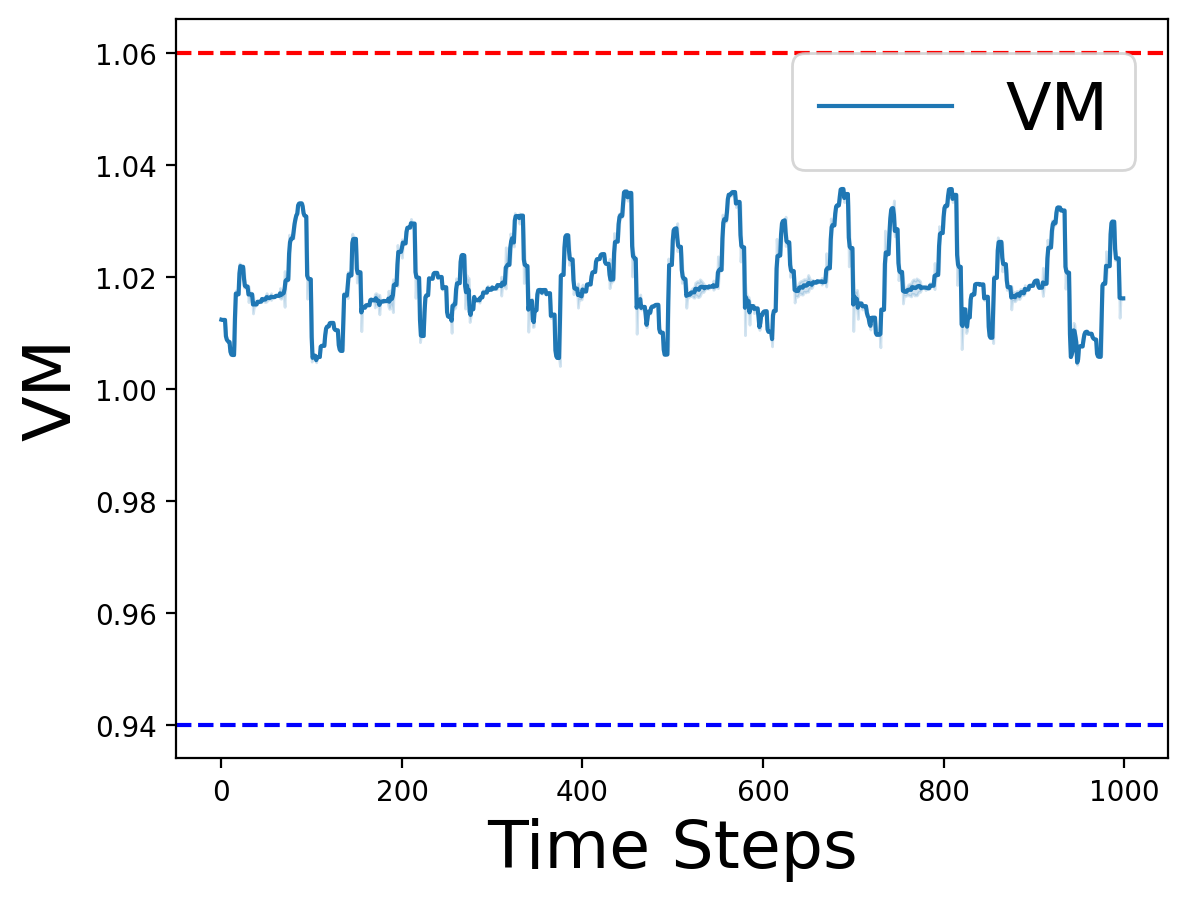}}
\caption{The testing curves of average voltage magnitudes by the  Constrained DRL in the IEEE 14-bus system.} \label{vmcurve14}
\end{minipage}
\hspace{-0.1cm}
\begin{minipage}{0.24\textwidth}
    \centering
    {\includegraphics[width=1\textwidth]{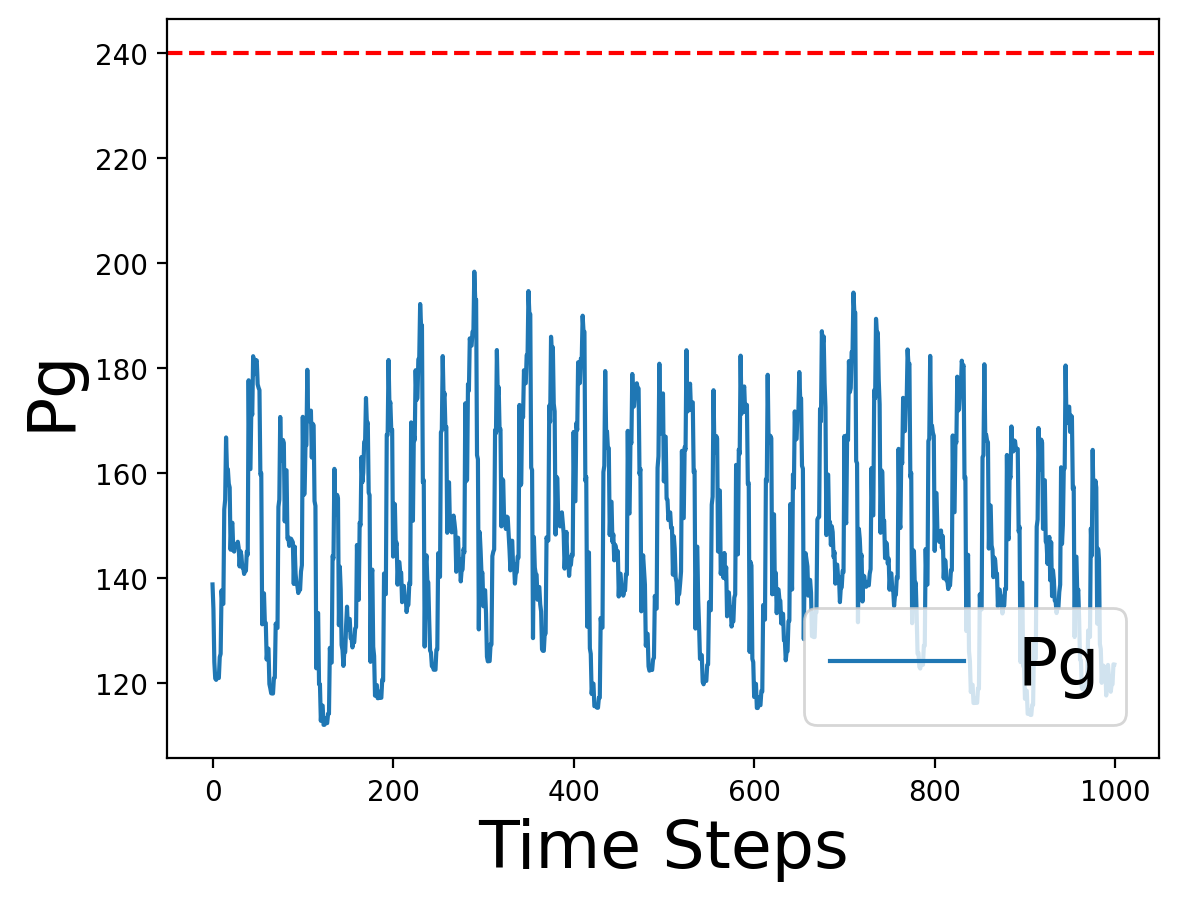}}
\caption{The testing curves of swing-bus generations by the  Constrained DRL in the IEEE 14-bus system.} \label{pgcurve14}
\end{minipage}
\vspace{-0.5cm}
\end{figure}

The reinforcement learning setting is defined as follows: the maximum number of iterations is 10000, the buffer size is 500, the discount factor for the reward is 0.99, the rate at which the target network updates is 0.005, the frequency of policy updates with delay is 2, and the frequency of dual updates is 500. We use three-layer feed-forward neural networks, each with 256 neurons, for both the critic and target networks. The actor and prediction networks consist of a spatio-temporal Chebyshev GCN and two-layer feed-forward neural networks with 256 neurons per layer. The activation functions used are rectified linear units (ReLU) for both the actor and critic networks, and a sigmoid unit only for the output of the actor network. Both network parameters are optimized using Adam with a learning rate of $10^{-3}$. The networks are trained after each time step using a mini-batch of 100 transitions, sampled uniformly from the replay buffer, which stores the entire history of transition tuples $(\bm{x}_{t-1}, \bm{A}_{t}, r_{t}, \bm{x}_{t})$.  Note that the actor networks have two independent GCN for the active and reactive power generation, respectively. This design can further boost the performance of the proposed CRL because it reduces the contradictory  actions by the active and reactive actions.

\subsection{Baseline Methods}
Before evaluating our proposed approach, we define several baseline methods for comparison.  First of all,  we consider two well-known DRL methods with Deep Q-Network (DQN) and Deep Deterministic Policy Gradient (DDPG). Secondly, we compare  the GCN-policy function with the fully connected neural networks (FNN), the convolutional neural networks (CNN) and Graph neural networks (GNN). Thirdly, we compare the proposed method with   the  optimization method knowing the future information. We also extend the  existing learning-based OPF methods, i.e., the  penalty method \cite{pan2022deepopf} and DC3 \cite{donti2020dc3}, to the reinforcement learning setting and compare them with the proposed algorithm.

\begin{figure}[htbp]   %
\centering
\begin{minipage}{0.24\textwidth}
    \centering
    {\includegraphics[width=1\textwidth]{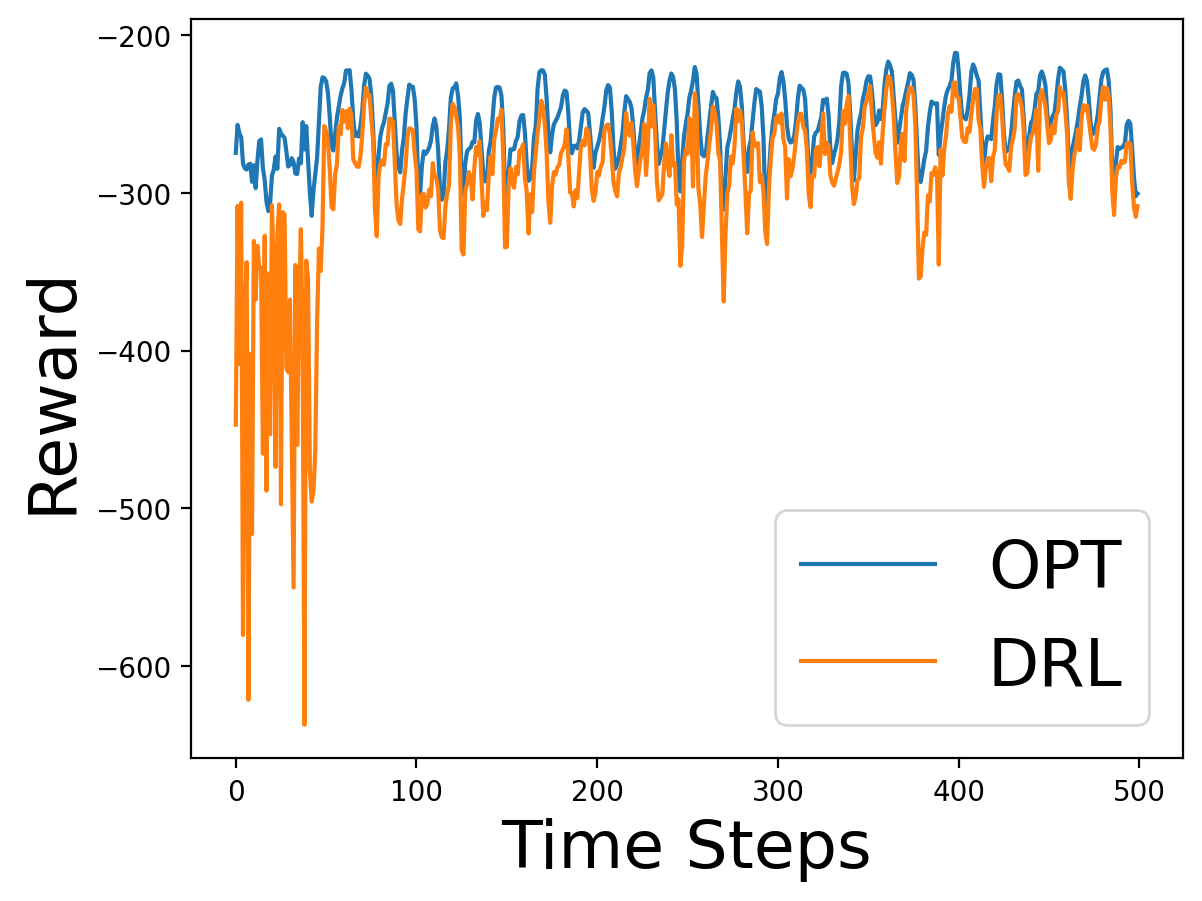}}
    \caption{The training curves of the  Constrained DRL and its corresponding optimal rewards by the optimization method in the IEEE 30-bus system.} \label{trainingcurve30bus}
\end{minipage}
\hspace{-0.1cm}
\begin{minipage}{0.24\textwidth}
    \centering
    {\includegraphics[width=1\textwidth]{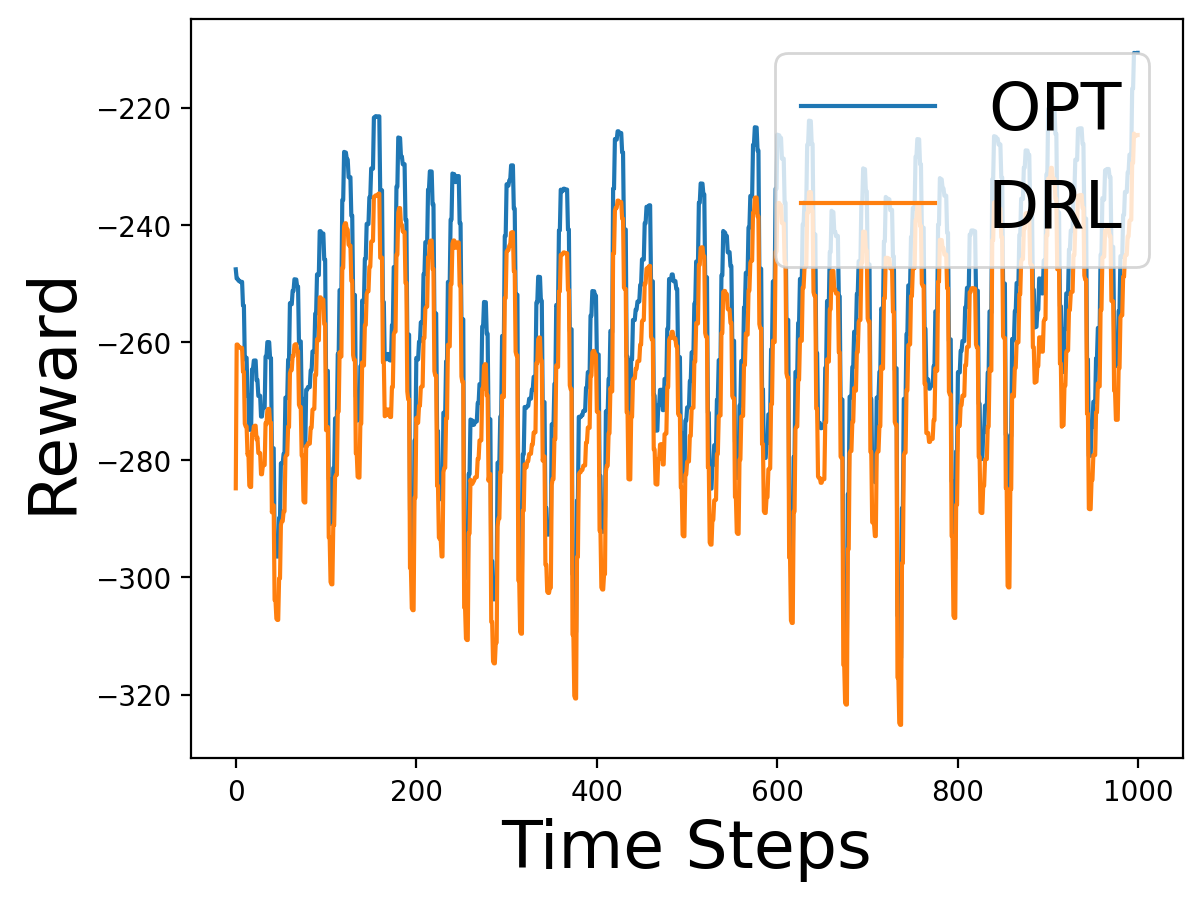}}
\caption{The testing curves of the  Constrained DRL and its corresponding optimal rewards by the optimization method in the IEEE 30-bus system.} \label{testingcurve30bus}
\end{minipage}
\hspace{-0.1cm}
\begin{minipage}{0.24\textwidth}
    \centering
    {\includegraphics[width=1\textwidth]{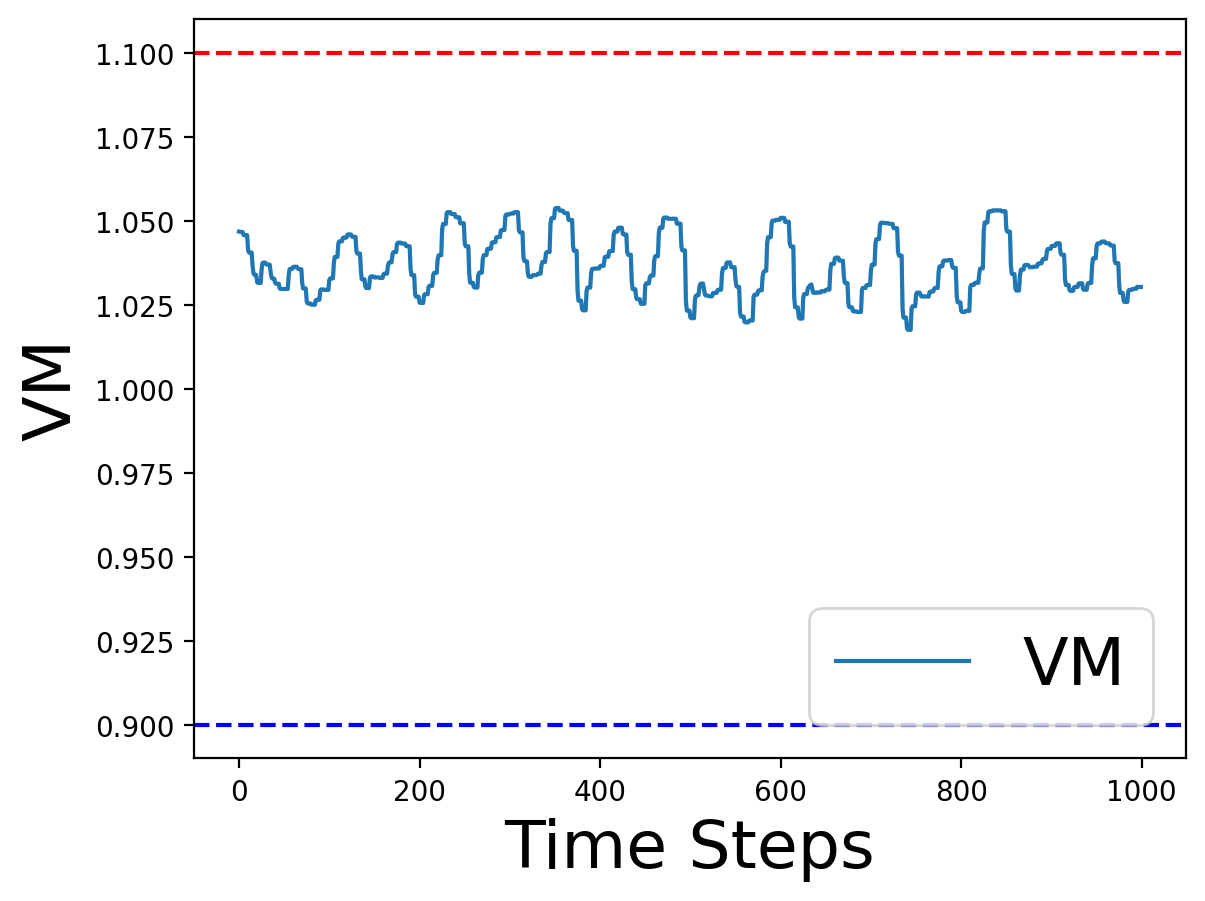}}
\caption{The testing curves of average voltage magnitudes by the  Constrained DRL in the IEEE 30-bus system.} \label{vmcurve30}
\end{minipage}
\hspace{-0.1cm}
\begin{minipage}{0.24\textwidth}
    \centering
    {\includegraphics[width=1\textwidth]{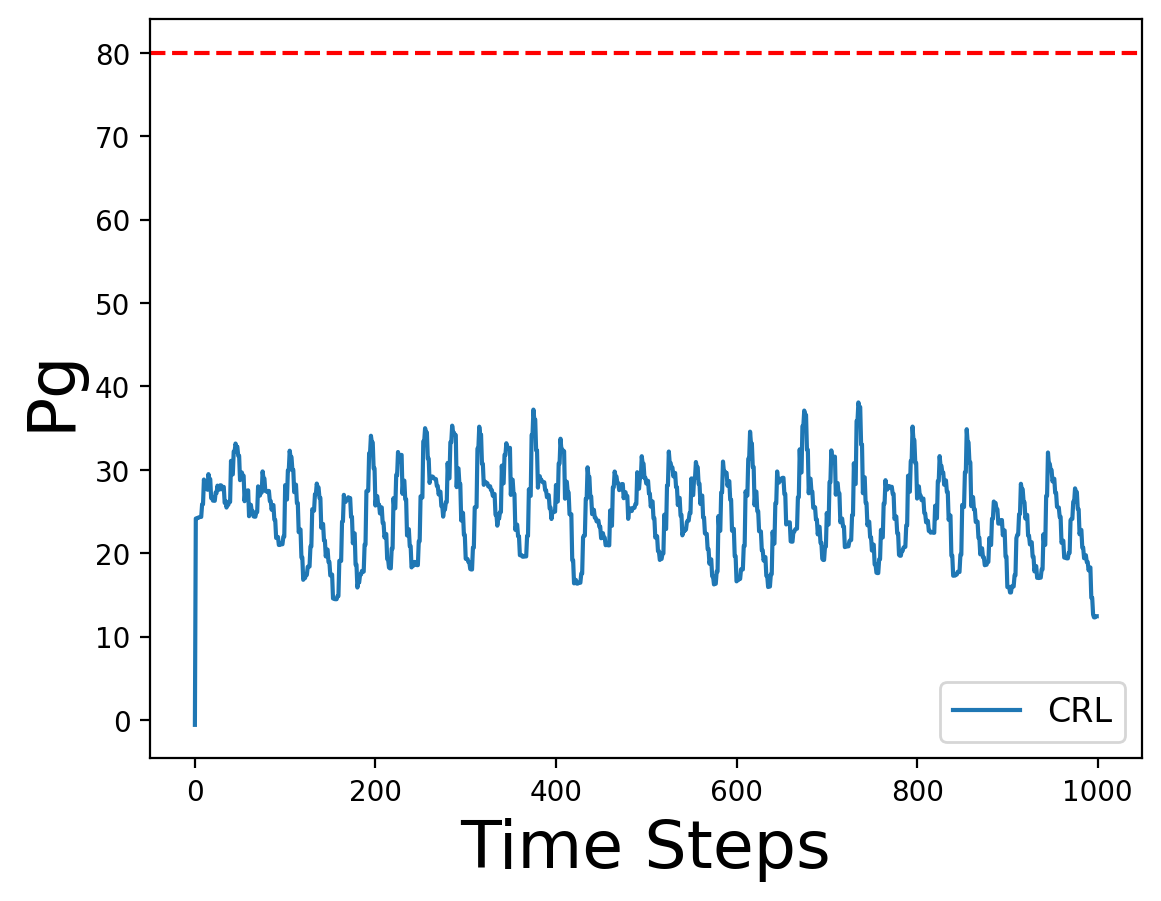}}
\caption{The testing curves of swing-bus generations by the  Constrained DRL in the IEEE 30-bus system.} \label{pgcurve30}
\end{minipage}
\vspace{-0.5cm}
\end{figure}

%
%
\subsection{Learning Curves and Optimal Curves}
\subsubsection{IEEE 14-bus System}
The learning curves in Fig. \ref{trainingcurve14bus} show the rewards, represented on the $y$-axis, at each time step on the $x$-axis. The exploration phase lasts 200 epochs, and our results are compared with an oracle solution that knows the future in the time horizon. The results indicate that our learning policy is very close to the optimal curves, demonstrating the effectiveness of our method in forecasting optimal actions without future information. The average gap between our DRL approach and the optimization method is only 2.52\%. After training, the policy function is tested on unseen scenarios, and the results, shown in Fig. \ref{testingcurve14bus}, show that the policy selects near-optimal actions with only 2.25\% optimal gaps. 

With the trained policy by the constrained DRL, we utilize this trained policy for testing new demands in the future 1000 samples to test the feasibility. Both the voltage magnitudes and power generations are feasible. The infeasible actions are likely to happen without the dual updates. For example, when ${g}_{i,t}, i\in \mathcal{G}_n$ of non-slack buses  predicted by the RL  is very small, ${g}_{i,t}, i\in \mathcal{G}_s$ of the slack bus will violate the upper bound constraint. Besides, the voltage magnitudes are easy to be violated if the reactive injections are too large or not enough.
Therefore,  in Fig. \ref{vmcurve14} and Fig. \ref{pgcurve14}, we show the swing-bus generation ${g}_{i,t}, i\in \mathcal{G}_s$   and the average voltage magnitude $ \frac{1}{N}\sum_{i\in \mathcal{N}} \abs{v_t}$, which shows the proposed CRL is always feasible.

\subsubsection{IEEE 30-bus System}
In Fig. \ref{trainingcurve30bus}, the learning curve for the proposed CRL in the IEEE 30-bus system with two BESSs at Bus 13 and Bus 22 is presented. The average optimal gap between the proposed algorithm and the optimization method with knowledge of future information in the time horizon is displayed for comparison and found to be only 3.39\%. The trained policy function is also tested on future scenarios, as shown in Fig. \ref{testingcurve30bus}, with results indicating that the policy selects near-optimal actions without any future information. Likewise, we also consider the feasibility of generation and voltage magnitudes in Figs. \ref{vmcurve30} and \ref{pgcurve30}. It also shows that the feasibility can be ensured for a large case.

\begin{figure}[htbp]   %
\centering
\begin{minipage}{0.24\textwidth}
    \centering
    {\includegraphics[width=1\textwidth]{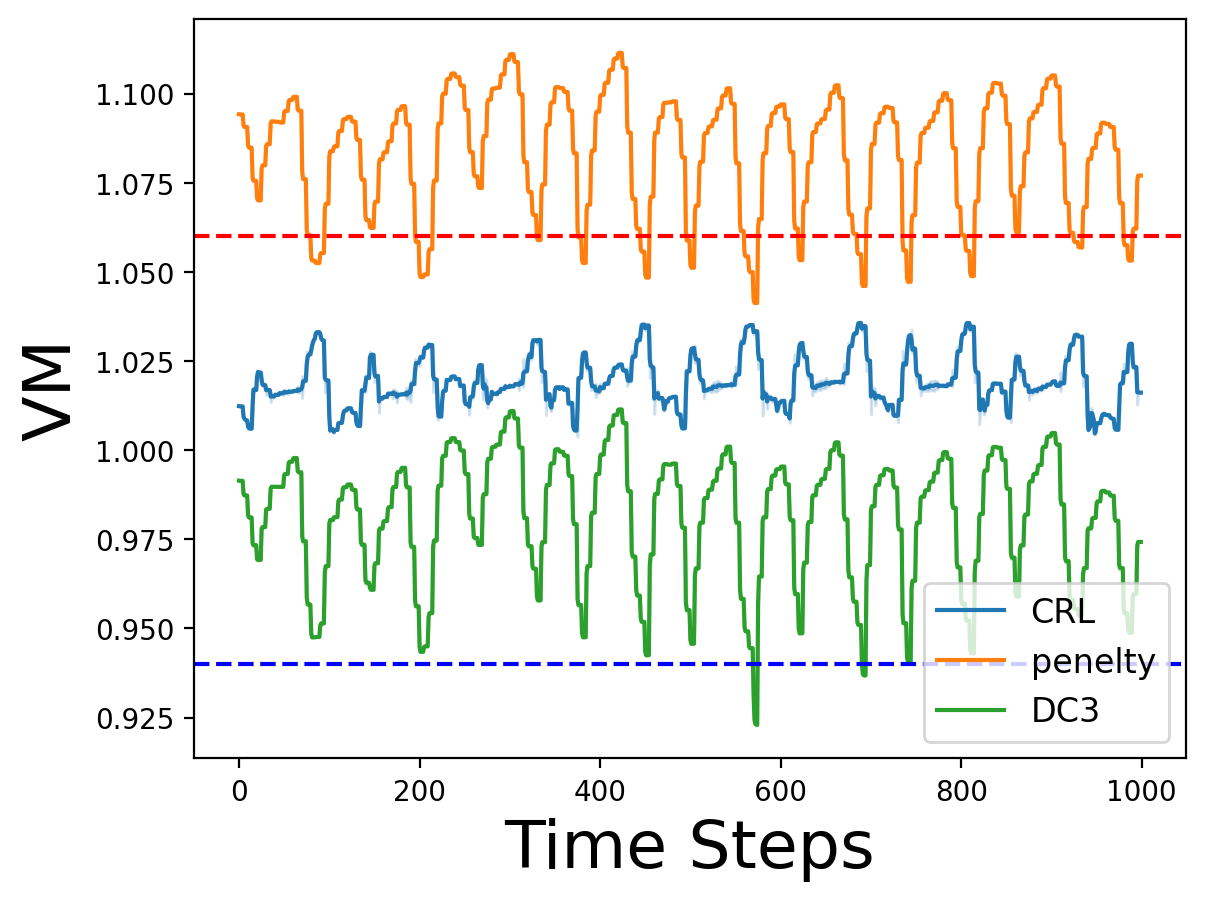}}
\caption{The testing curves of average voltage magnitudes by the penalty method, the DC3 method and the proposed CRL in the IEEE 14-bus system.} \label{VM_cst}
\end{minipage}
\hspace{-0.1cm}
\begin{minipage}{0.24\textwidth}
    \centering
    {\includegraphics[width=1\textwidth]{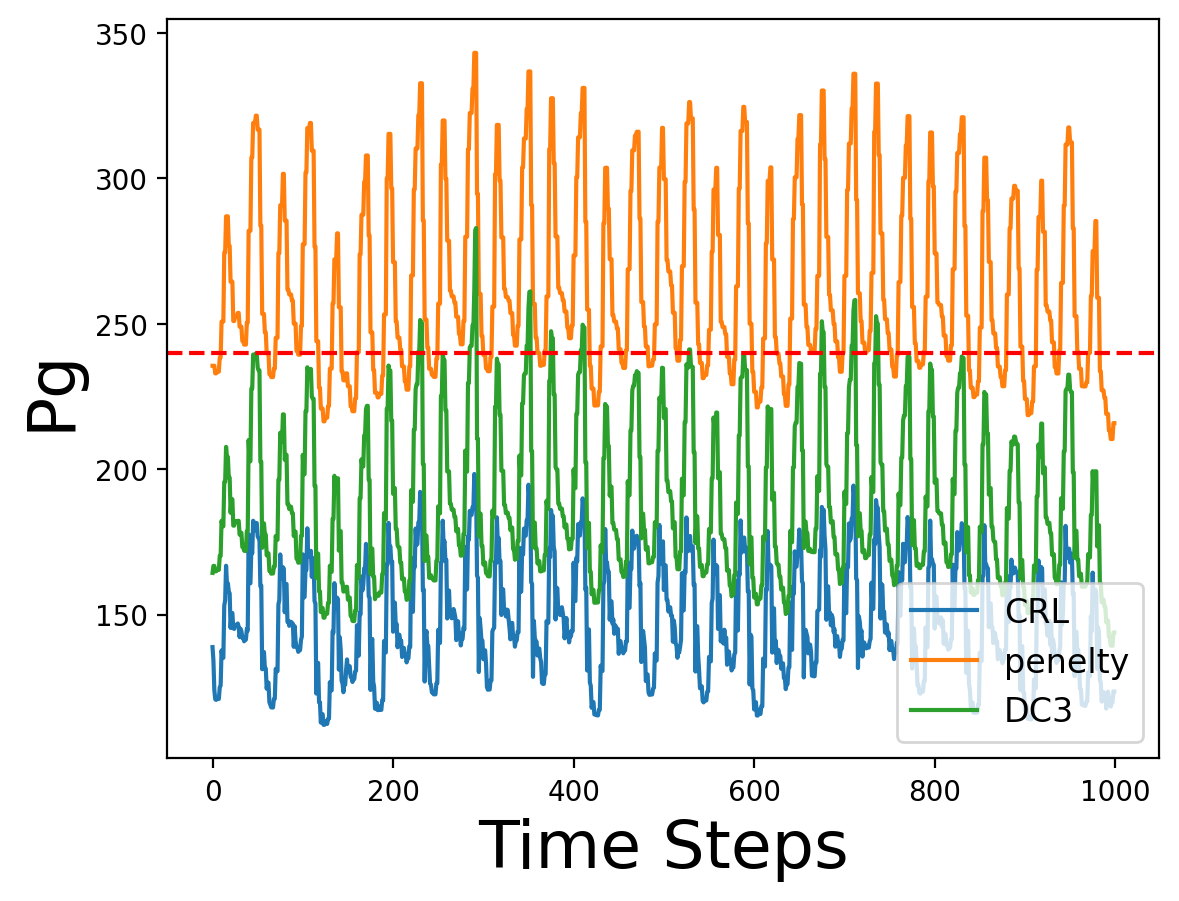}}
\caption{The testing curves of swing-bus generations by the penalty method, the DC3 method and the proposed CRL in the IEEE 14-bus system.} \label{Pg_cst}
\end{minipage}
\vspace{-0.5cm}
\end{figure}
\subsection{Feasibility Comparison}
We compared the proposed primal-and-dual CRL with two baselines, which includes     the  penalty method \cite{pan2022deepopf} and DC3 \cite{donti2020dc3}. Typically, the  penalty method gives the violation penalty as a  rectified linear unit function, which are considered in the reward. DC3 methods consider the equality and inequality constraints as $\norm{\cdot}_2^2$, which are considered in the policy objectives without the primal-and-dual update process.
 Figs. \ref{VM_cst} and \ref{Pg_cst}  shows that the proposed constrained DRL can ensure 100\% feasibility, whereas the traditional DRL has only 80.78\% feasibility rate. The penalty method provides 13.50\% feasibility rate and the DC3 method has 96.17\% feasibility rate for the voltage magnitudes, and 29.46\% feasibility rate and  98.56\% feasibility rate for the swing-bus generations. This indicates that the key element to enforce the 100\% feasibility is the dual update.

\begin{figure}[htbp]   %
\centering
\begin{minipage}{0.24\textwidth}
    \centering
    {\includegraphics[width=1\textwidth]{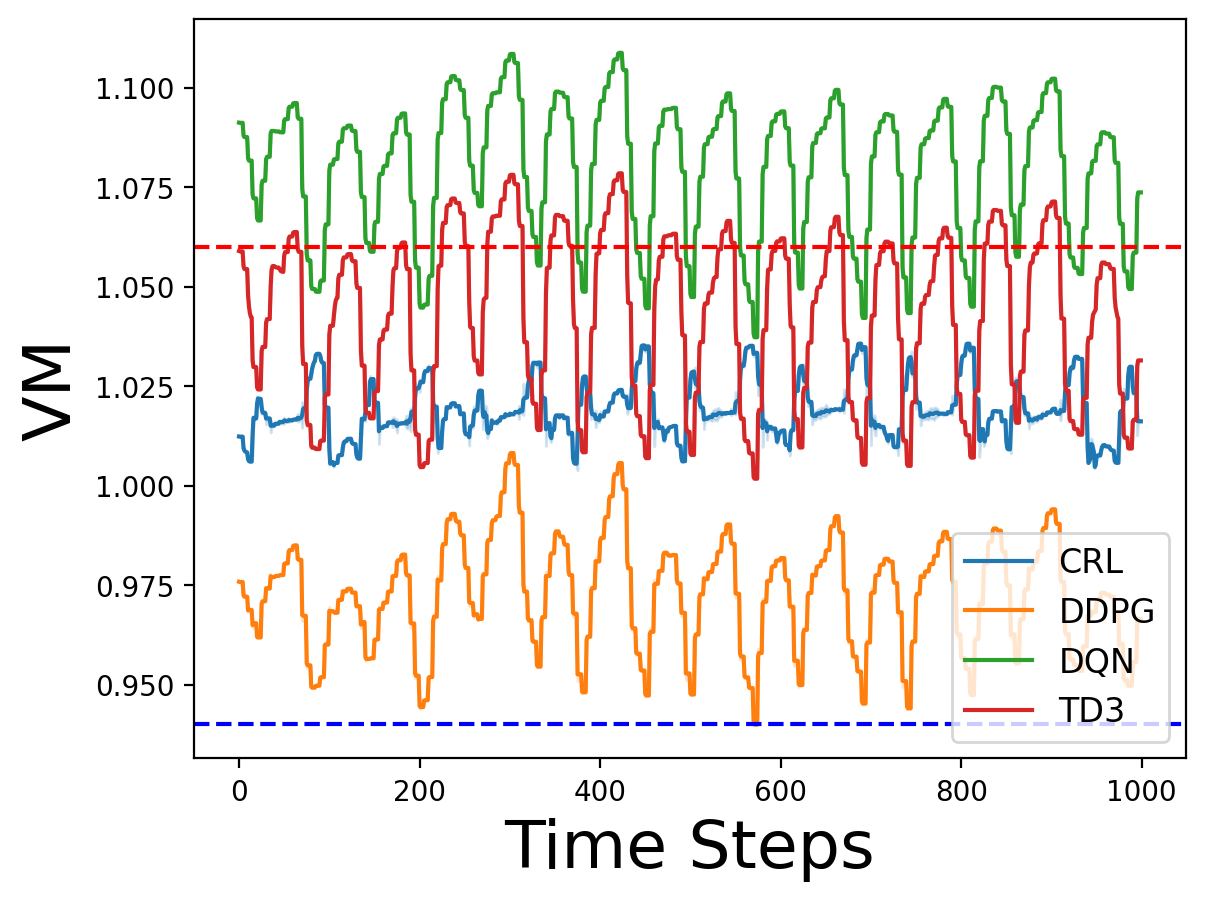}}
    \caption{The testing curves of average voltage magnitudes by DDPG, DQN, TD3 and the proposed CRL in the IEEE 14-bus system.} \label{trainingcurve_DRL}
\end{minipage}
\hspace{-0.1cm}
\begin{minipage}{0.24\textwidth}
    \centering
    {\includegraphics[width=1\textwidth]{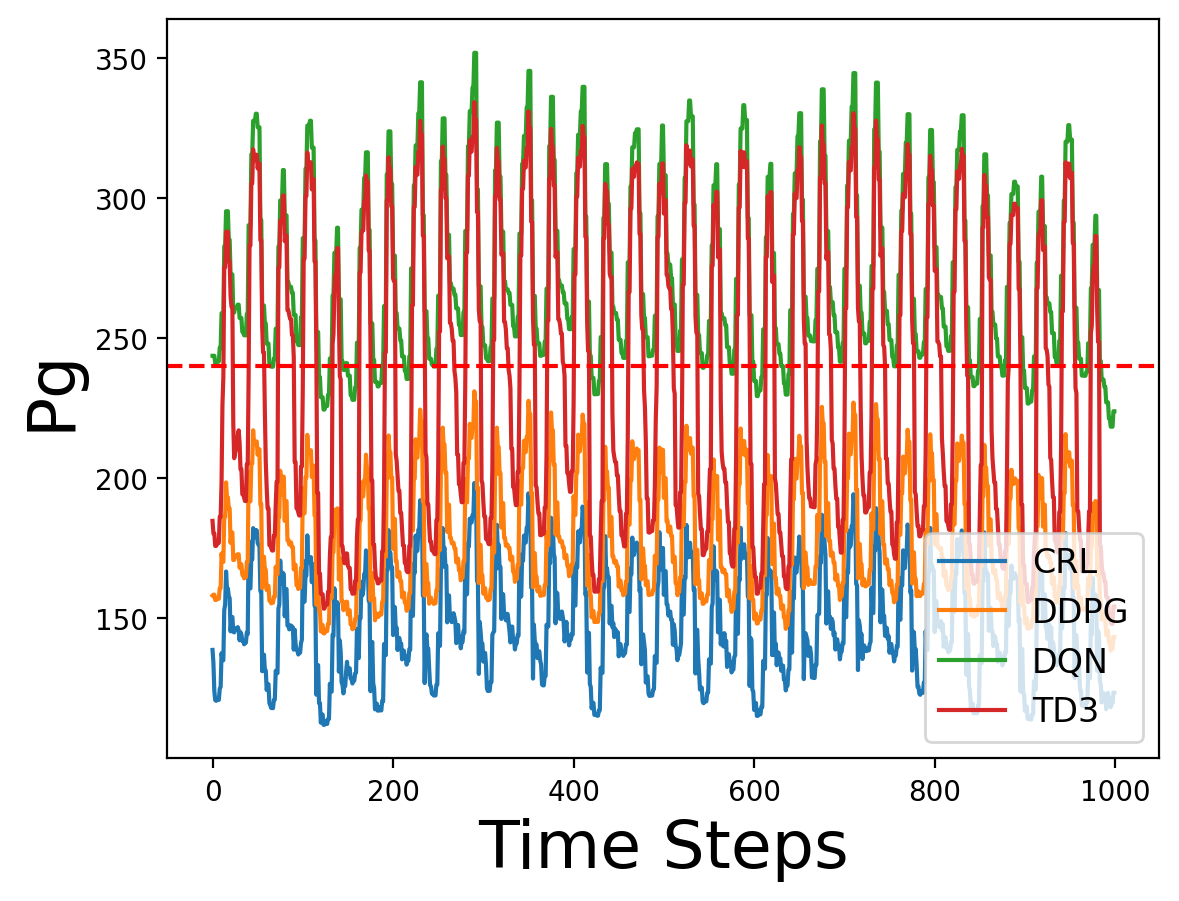}}
\caption{The testing curves of swing-bus generations by DDPG, DQN, TD3 and the proposed CRL in the IEEE 14-bus system.} \label{testingcurve_DRL}
\end{minipage}
\vspace{-0.5cm}
\end{figure}

\subsection{Comparison of Reinforcement Learning}
We also compare the proposed CRL with the existing deep reinforcement learning methods, i.e., DQN, DDPG and TD3 without primal-and-dual processes, based on the IEEE-14 bus system. The proposed CRL method has better   testing rewards, i.e., 2.25\%, where TD3 converges into 5.04\%, DQN converges into 7.33\% and DDPG converges into 8.10\%.
We show the testing voltage magnitude curves and testing power generation curves in Figs. \ref{trainingcurve_DRL} and \ref{testingcurve_DRL}.  In particular, TD3, DQN and DDPG have 52.39\%, 14.95\% and 100\% power generation feasibility rate, and 70.89\%, 17.72\% and 99.27\% voltage magnitude feasibility rate, while the proposed CRL has both 100\% power generation feasibility rate and 100\% voltage magnitude feasibility rate. Besides, the proposed method can control  voltage magnitude profiles with less variations, which are closed to 1 p.u.

\begin{figure}[htbp]   %
\centering
\begin{minipage}{0.24\textwidth}
    \centering
    {\includegraphics[width=1\textwidth]{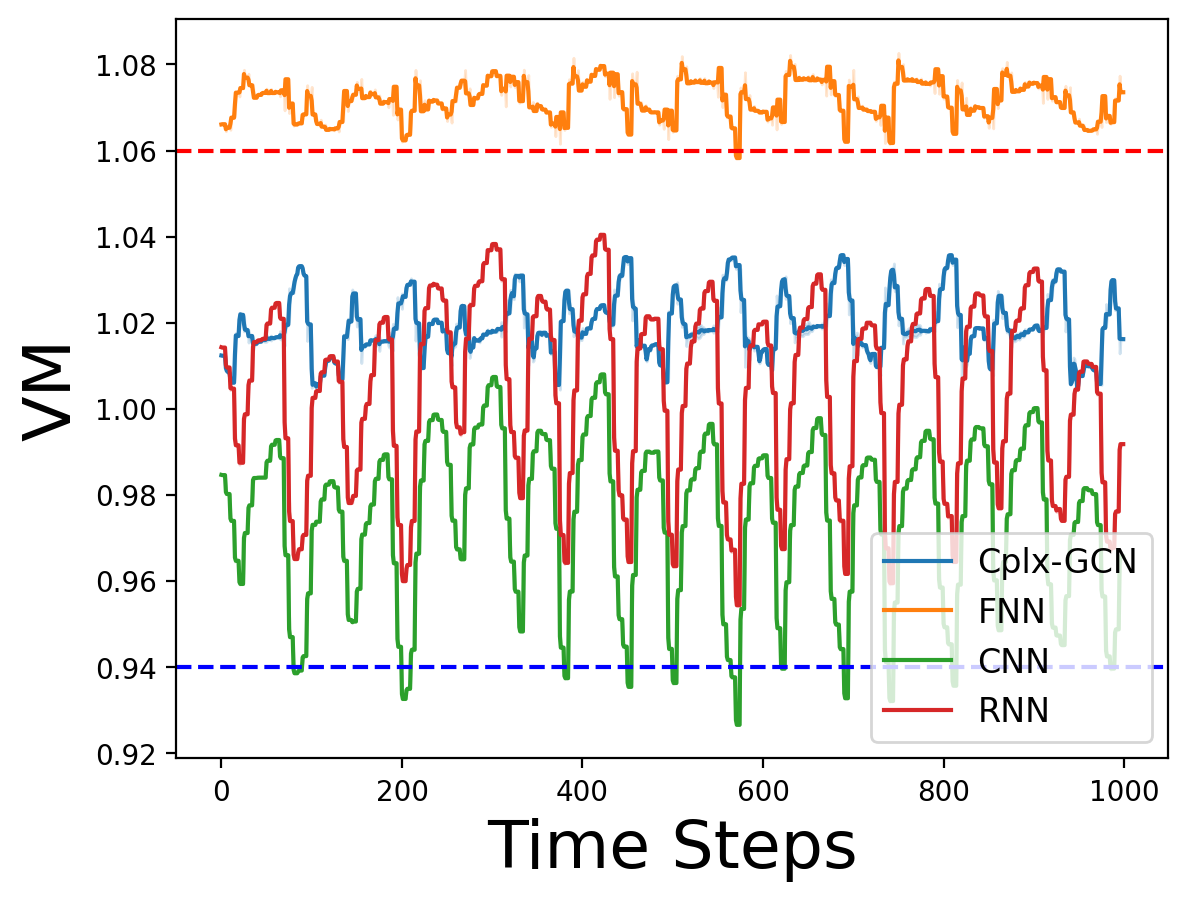}}
    \caption{The testing curves of average voltage magnitudes by FNN, CNN, RNN and Cplx-GCN policies in the IEEE 14-bus system.} \label{trainingcurve_NN}
\end{minipage}
\hspace{-0.1cm}
\begin{minipage}{0.24\textwidth}
    \centering
    {\includegraphics[width=1\textwidth]{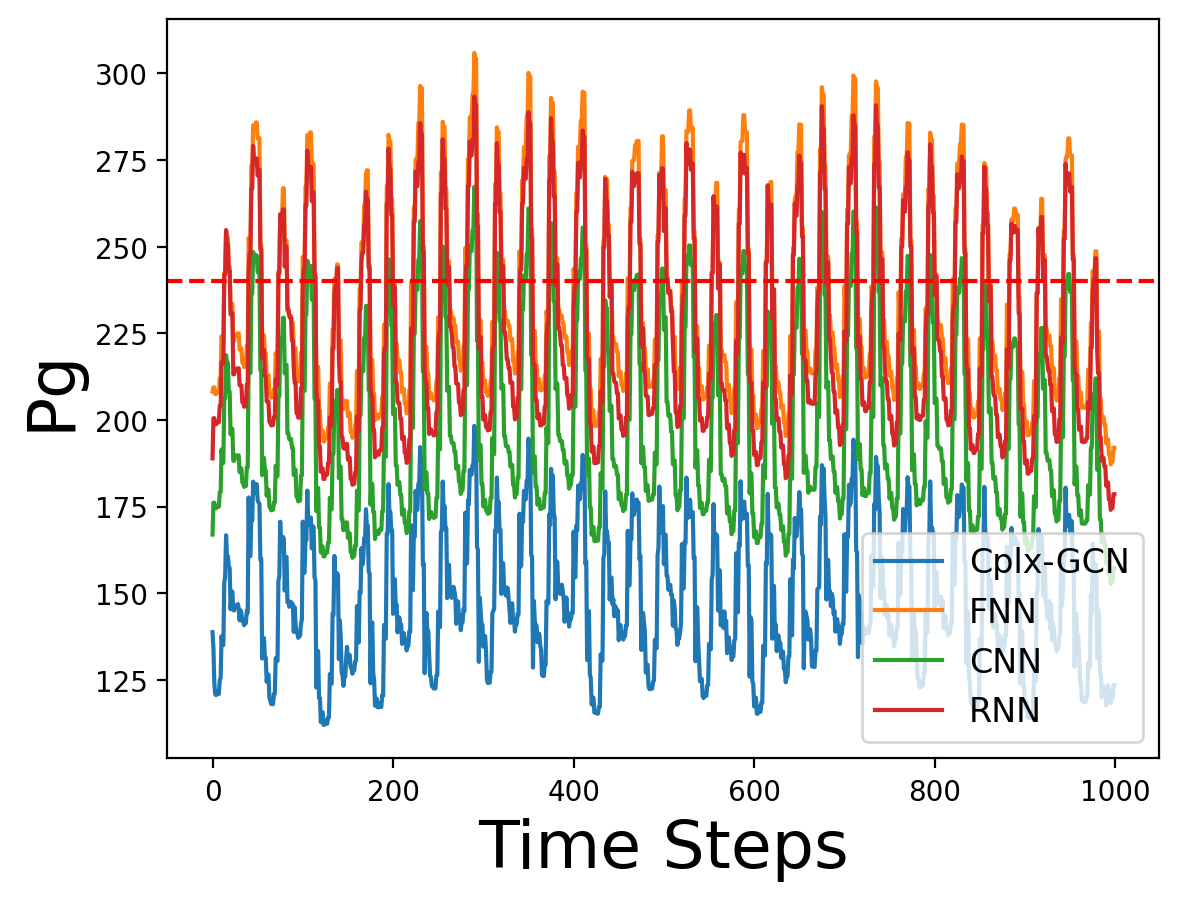}}
\caption{The testing curves of swing-bus generations by FNN, CNN, RNN and Cplx-GCN policies in the IEEE 14-bus system.} \label{testingcurve_NN}
\end{minipage}
\hspace{-0.4cm}
\end{figure}

\subsection{Comparison of Policy Neural Networks}
Moreover,  we compare the Cplx-STGCN policy neural networks with different policy neural networks, i.e. FNN, CNN, and RNN. Similarly, the testing average optimal gaps of FNN, CNN and RNN are 4.99\%,   which illustrates the advantage of Cplx-STGCN (i.e. 2.25\%) to extract the spatio-temporal features for the policy function. In Fig. \ref{trainingcurve_NN} and \ref{testingcurve_NN},   FNN, CNN and RNN have    65.41\%, 89.45\% and 63.37\% power generation feasibility rate, and 1.09\%, 92.60\% and 100\% voltage magnitude feasibility rate. In contrast, the proposed CRL are always feasible.


\section{Conclusion}
In this paper, we proposed a novel constrained reinforcement learning   using the prime-dual decomposition to update the policy networks, in conjunction with a double-Q learning method to update the critic networks for solving a multi-stage SDOPF. We further proved the convergence of the proposed CRL under mild assumptions. 
The numerical results show that the actions chosen with the policy derived are close to the optimal actions of an oracle OPF that knows the future,   and are feasible 100\% of the times. Compared with other RL methods, the proposed CRL achieves the higher reward. Compared with other neural networks, the proposed Cplx-STGCN policy function has better performance in extracting the spatio-temporal features for voltage phasors. Besides, the feasible rate of the proposed CRL is higher than other constrained neural network methodologies.

\appendix
The proof of Lemma 1 is provided as follows.

\emph{Proof:} Since $(\phi^*, \bm{\lambda}^*, \bm{\mu}^*)$ is a saddle point for unaugmented Lagrangian$\mathcal{L}_\phi$, we have
\begin{equation}
\begin{aligned}
 \mathcal{L}_\phi (\phi^*, \bm{\lambda}^*, \bm{\mu}^*) \le  \mathcal{L}_\phi (\phi^{k+1}, \bm{\lambda}^*, \bm{\mu}^*)
\end{aligned}
\end{equation} 
 Using $ \bm{L} \pi_{\phi} (\bm{x}) - \bm{b}= \bm{0}$ and $  [\bm{K} \pi_{\phi} (\bm{x})  -  \bm{c}]_+  = \bm{0} $, the left sides is $p^* = -Q^*(\bm{x}, \pi_{\phi^{*}}(\bm{x})) $. With $p^{k+1} = -Q^*(\bm{x}, \pi_{\phi^{k+1}}(\bm{x}))$, this can be written as
 \begin{equation}
\begin{aligned}
p^* \le p^{k+1} +    \bm{\lambda}^{*\top}  [\bm{L} \pi_{\phi} (\bm{x}) - \bm{b}]    + \bm{\mu}^{*\top}   [\bm{K} \pi_{\phi} (\bm{x})  -  \bm{c}]_+
\end{aligned}
\end{equation} 
We can conclude the first key inequality:
 \begin{equation}
\begin{aligned}\label{key1}
p^* \le p^{k+1} +   \bm{\lambda}^{*\top}  \bm{r}^{k+1}_\lambda + \bm{\mu}^{*\top}  \bm{r}^{k+1}_\mu 
\end{aligned}
\end{equation}

By definition, $\phi^{k+1}$ minimizes $\mathcal{L}_\phi^\alpha (\phi, \bm{\lambda}^k, \bm{\mu}^k)$.  The optimality condition is 
 \begin{equation}
\begin{aligned}
0 \in  & \partial (-Q^*(\pi_{\phi^{k+1}}(\bm{x}))  \partial   \pi_{\phi^{k+1}}(\bm{x})  + \bm{L}^\top \bm{\lambda}^k  \partial   \pi_{\phi^{k+1}}(\bm{x}) +     \alpha_\lambda \bm{L}^\top    \\
&   (\bm{L}  \pi_{\phi^{k+1}}(\bm{x}) - \bm{b})   \partial   \pi_{\phi^{k+1}}(\bm{x}) + \bm{K}^\top  \bm{D}^{(k+1)} \bm{\mu}^k  \partial   \pi_{\phi^{k+1}}(\bm{x}) \\
& +  \alpha_\mu \bm{K}^\top  \bm{D}^{(k+1)} (\bm{K}  \pi_{\phi^{k+1}}(\bm{x}) - \bm{c})_+  \partial   \pi_{\phi^{k+1}}(\bm{x})\\
= & \partial (-Q^*(\pi_{\phi^{k+1}}(\bm{x}))  \partial   \pi_{\phi^{k+1}}(\bm{x})  + \bm{L}^\top \Big(  \bm{\lambda}^k   +   \alpha_\lambda  (\bm{L}  \pi_{\phi^{k+1}}(\bm{x})     \\
&      - \bm{b})  \Big) \partial   \pi_{\phi^{k+1}}(\bm{x}) + \bm{K}^\top  \bm{D}^{(k+1)} \Big( \bm{\mu}^k +  \alpha_\mu   (\bm{K}  \pi_{\phi^{k+1}}(\bm{x})    \\
&   - \bm{c})_+  \Big)  \partial   \pi_{\phi^{k+1}}(\bm{x})\\
& = \partial (-Q^*(\pi_{\phi^{k+1}}(\bm{x}))  \partial   \pi_{\phi^{k+1}}(\bm{x})  + \bm{L}^\top  \bm{\lambda}^{k+1} \partial   \pi_{\phi^{k+1}}(\bm{x})   \\
& +  \bm{K}^\top  \bm{D}^{(k+1)}    \bm{\mu}^{k+1} \partial   \pi_{\phi^{k+1}}(\bm{x})
\end{aligned}
\end{equation} 
where $\bm{D}^{(k+1)}$ is a diagonal matrix
 \begin{equation}
\begin{aligned}
\bm{D}^{(k+1)}_{ii}  = \begin{cases}
      1 & [\bm{K}]_i^\top[ \pi_{\phi^{k+1}}(\bm{x}) ]_i > \bm{c}_i \\
      0 &  [\bm{K}]_i^\top[ \pi_{\phi^{k+1}}(\bm{x}) ]_i \le \bm{c}_i 
    \end{cases} 
\end{aligned}
\end{equation} 

This implies that $\phi^{k+1}$ minimizes 
 \begin{equation}
\begin{aligned}
(-Q^*(\pi_{\phi }(\bm{x})) +    ( \bm{\lambda}^{k+1})^\top  \bm{L} 
 \pi_{\phi }(\bm{x})+  (\bm{\mu}^{k+1})^\top  \bm{D}^{(k+1)}  \bm{K}  \pi_{\phi }(\bm{x}) 
\end{aligned}
\end{equation} 
It follows
 \begin{equation}
\begin{aligned}
& (-Q^*(\pi_{\phi^{k+1} }(\bm{x}))) +     (\bm{\lambda}^{k+1})^\top  \bm{L} 
 \pi_{\phi^{k+1} }(\bm{x})+   \\
&  (\bm{\mu}^{k+1})^\top  \bm{D}^{(k+1)}  \bm{K}  \pi_{\phi^{k+1} }(\bm{x}) \le (-Q^*(\pi_{\phi^* }(\bm{x}))) +          \\
& \bm{\lambda}^{k+1}  \bm{L}  \pi_{\phi^* }(\bm{x})+ \bm{\mu}^{k+1}  \bm{D}^{(k+1)}  \bm{K}  \pi_{\phi^* }(\bm{x}) 
\end{aligned}
\end{equation} 
Using $\bm{L}  \pi_{\phi^* }(\bm{x}) = \bm{b}$ and $\bm{K}  \pi_{\phi^* }(\bm{x}) \preceq \bm{c}$, we can obtain the second key inequation:
 \begin{equation}
\begin{aligned}\label{key2}
&p^{k+1} - p^* \\
&\le - (\bm{\lambda}^{k+1})^\top \bm{r}^{k+1}_\lambda - (\bm{\mu}^{k+1})^\top   \bm{D}^{(k+1)}  \bm{K} (\pi_{\phi^{k+1} }(\bm{x}) - \pi_{\phi^* }(\bm{x})  ) \\
&  \le - (\bm{\lambda}^{k+1})^\top \bm{r}^{k+1}_\lambda - (\bm{\mu}^{k+1})^\top \bm{D}^{(k+1)}  (\bm{K} \pi_{\phi^{k+1} }(\bm{x}) - \bm{c}  ) \\
& = - (\bm{\lambda}^{k+1})^\top \bm{r}^{k+1}_\lambda - (\bm{\mu}^{k+1})^\top     (\bm{K}\pi_{\phi^{k+1} }(\bm{x}) - \bm{c}  )_+ \\ 
& = - (\bm{\lambda}^{k+1})^\top \bm{r}^{k+1}_\lambda - (\bm{\mu}^{k+1})^\top \bm{r}^{k+1}_\mu 
\end{aligned}
\end{equation} 
Adding \eqref{key1} and \eqref{key2}, regrouping terms, and multiplying through by 2 gives
 \begin{equation}
\begin{aligned} 
    2 (\bm{\lambda}^{k+1}   -  \bm{\lambda}^{*} )^\top \bm{r}^{k+1}_\lambda +   2 (\bm{\mu}^{k+1}  -  \bm{\mu}^{*})^\top \bm{r}^{k+1}_\mu \le 0
\end{aligned}
\end{equation} 
We begin by rewriting the first term. Substituting  $\bm{\lambda}^{k+1} = \bm{\lambda}^{k} + \alpha_\lambda \bm{r}^{k+1}_\lambda $ and $\bm{\mu}^{k+1} = \bm{\mu}^{k} + \alpha_\mu \bm{r}^{k+1}_\mu $:
 \begin{equation}
\begin{aligned} 
  &  2 (\bm{\lambda}^{k}  -  \bm{\lambda}^{*} )^\top \bm{r}^{k+1}_\lambda + \alpha_\lambda \norm{\bm{r}^{k+1}_\lambda}_2^2 + \alpha_\lambda \norm{\bm{r}^{k+1}_\lambda}_2^2 +    \\
  &  2 (\bm{\mu}^{k}  -  \bm{\mu}^{*} )^\top \bm{r}^{k+1}_\mu + \alpha_\mu \norm{\bm{r}^{k+1}_\mu}_2^2 + \alpha_\mu \norm{\bm{r}^{k+1}_\mu}_2^2
\end{aligned}
\end{equation} 
and substituting $ \bm{r}^{k+1}_\lambda = \frac{1}{\alpha_\lambda} (\bm{\lambda}^{k+1}  -  \bm{\lambda}^{k})$ and $ \bm{r}^{k+1}_\mu = \frac{1}{\alpha_\mu} (\bm{\mu}^{k+1}  -  \bm{\mu}^{k})$ in the first two terms gives
 \begin{equation}
\begin{aligned} 
  &  \frac{2}{\alpha_\lambda} (\bm{\lambda }^{k}  -  \bm{\lambda}^{*})^\top (\bm{\lambda}^{k+1}  -  \bm{\lambda}^{k}) + \frac{1}{\alpha_\lambda} \norm{\bm{\lambda}^{k+1}  -  \bm{\lambda}^{k} }_2^2 \\
  &  + \alpha_\lambda \norm{\bm{r}^{k+1}_\lambda}_2^2 + \frac{2}{\alpha_\mu} (\bm{\mu }^{k}  -  \bm{\mu}^{*})^\top (\bm{\mu}^{k+1}  -  \bm{\mu}^{k})\\
  & + \frac{1}{\alpha_\mu} \norm{\bm{\mu}^{k+1}  -  \bm{\mu}^{k} }_2^2 + \alpha_\mu \norm{\bm{r}^{k+1}_\mu}_2^2
\end{aligned}
\end{equation} 
Since $\bm{\lambda}^{k+1} -  \bm{\lambda}^{k} = (\bm{\lambda}^{k+1} -  \bm{\lambda}^{*}) - (\bm{\lambda}^{k} -  \bm{\lambda}^{*}) $ and $\bm{\mu}^{k+1} -  \bm{\mu}^{k} = (\bm{\mu}^{k+1} -  \bm{\mu}^{*}) - (\bm{\mu}^{k} -  \bm{\mu}^{*}) $, this can be written as
 \begin{equation}
\begin{aligned} 
   &\frac{1}{\alpha_\lambda} (\norm{\bm{\lambda}^{k+1}  -  \bm{\lambda}^{*} }_2^2  - \norm{\bm{\lambda}^{k}  -  \bm{\lambda}^{*} }_2^2) + \alpha_\lambda \norm{\bm{r}^{k+1}_\lambda}_2^2 + \\
  & \frac{1}{\alpha_\mu} (\norm{\bm{\mu}^{k+1}  -  \bm{\mu}^{*} }_2^2  - \norm{\bm{\mu}^{k}  -  \bm{\mu}^{*} }_2^2) + \alpha_\mu \norm{\bm{r}^{k+1}_\mu}_2^2.
\end{aligned}
\end{equation} 
Therefore, we can obtain 
 \begin{equation}
\begin{aligned} 
   V^{k+1}  \le  V^{k}  - \alpha_\lambda \norm{\bm{r}^{k+1}_\lambda}_2^2 - \alpha_\mu \norm{\bm{r}^{k+1}_\mu}_2^2
\end{aligned}
\end{equation} 
This  states that $V^{k}$ decreases in each iteration by an amount that only depends on the norm of the residual. Because $V^{k}\le V^{0}$, it follows that $\bm{\lambda}^k$ and $\bm{\mu}^k$ are bounded. Iterating the inequality above gives that 
 \begin{equation}
\begin{aligned} 
    \alpha_\lambda \sum_{k=0}^{\infty} \norm{\bm{r}^{k+1}_\lambda}_2^2 +   \alpha_\mu \sum_{k=0}^{\infty} \norm{\bm{r}^{k+1}_\mu}_2^2 \le V^0
\end{aligned}
\end{equation} 
which means that $\bm{r}^{k+1}_\lambda \rightarrow 0$ and $\bm{r}^{k+1}_\mu \rightarrow 0$ as $k\rightarrow 0$. 

 This completes the proof.
\hfill $\blacksquare$

Together with Assumption A.1 and Lemma 1, both the primal and dual variables converge into a saddle point $(\phi^*, \bm{\lambda}^*, \bm{\mu}^*)$.

\begin{footnotesize}

\end{footnotesize}

\end{document}